%% file: ma5-korea-report.tex
\documentclass[hyperref,11pt]{cernyrep}
\pdfoutput=1
\usepackage{hyperref,url,xcolor,cite,color}
\usepackage[utf8]{inputenc}
\hypersetup{bookmarks=true,unicode=true,pdftoolbar=true,pdfmenubar=true,
  pdffitwindow=false,pdfstartview={FitH},
  pdfnewwindow=true,colorlinks=true,
  linkcolor=blue,citecolor=magenta,filecolor=magenta,urlcolor=cyan}

\input{0-commands.tex}

\begin{document}

\setcounter{tocdepth}{0}
\thispagestyle{empty}

\vspace{1cm}

\begin{center}
{\Large {\bf PROCEEDINGS OF THE FIRST MADANALYSIS 5 WORKSHOP\\[4mm]}}
{\Large {\bf  ON LHC RECASTING IN KOREA}}
\end{center}

\vspace{0.1cm}
\input {0-authors.tex}
\vspace{1cm}

\begin{center}
  {\large {\bf Abstract}}\\[.2cm]
\end{center}
We present the activities performed during the first \ma\ workshop on LHC
recasting that has been organized at High 1 (Gangwon privince, Korea) on August
20-27, 2017. This report includes details on the implementation in the \ma\
framework of eight ATLAS and CMS analyses, as well as a description of the
corresponding validation and the various issues that have been observed.
\vspace{1cm}
\begin{center}
{\bf Acknowledgements}\\[.2cm]
\end{center}
We are grateful to the local staff (Eunbi Jang, Sunmi Wee, Jieun Jeong and Brad
Kwon) who contributed a lot to the stimulating and lively atmosphere in which we
have worked, and to Nicolas Bizot, Giacomo Cacciapaglia, Valentin Hirschi,
Pyungwon Ko and Hwi-Dong Yoo for their nice lectures.
It has been made possible to organize this event thanks to the
amazing support of Global Research Funding project in Hanyang University, the
National Research Foundation of Korea (NRF) for grant
funded by the Korea government (MEST) (contracts NRF-2015R1A2A1A15052408 and
NRF-2017R1A2B4002498), of KIAS and of the France Korea Particle Physics and
e-science Laboratory (FKPPL) of the CNRS.
\newpage

\vspace{1cm}

\thispagestyle{empty}
\setcounter{page}{2}

\begin{center}
\input{0-addresses.tex}
\end{center}

\newpage

\tableofcontentscern

\newpage
\thispagestyle{empty}
$~$\newpage
\setcounter{page}{5}

\setcounter{figure}{0}
\setcounter{table}{0}
\setcounter{section}{0}
\setcounter{equation}{0}
\setcounter{footnote}{0}
\clearpage

\input{1-intro.tex}
\AddToContent{S.~Bein, G.~Chalons, E.~Conte, B.~Fuks, T.~Kim, S.J.~Lee, D.~Sengupta and J.~Sonneveld}
\newpage

\superpart{Dark Matter}
\newpage
\thispagestyle{empty}
$~$\newpage
\setcounter{page}{9}

\input{8-atlas_conf_2016_086.tex}
\AddToContent{B.~Fuks, M.~Zumbihl}
\renewcommand{\thesection}{\arabic{section}}

\input{7-atlas_1707_01302}
\AddToContent{S.~Jeon, Y.~Kang, G.~Lee, C.~Yu}
\renewcommand{\thesection}{\arabic{section}}

\input{9-atlas_1711_03301}
\AddToContent{D.~Sengupta}
\renewcommand{\thesection}{\arabic{section}}
\newpage
\thispagestyle{empty}

\input{3-atlas_1704_03848}

\AddToContent{S.~Baek and T.~H.~Jung}
\renewcommand{\thesection}{\arabic{section}}
\newpage
\thispagestyle{empty}
$~$\newpage

\input{2-cms_exo_16_012.tex}
\AddToContent{S.~Ahn, J.~Park and W.~Zhang}
\renewcommand{\thesection}{\arabic{section}}
\newpage
\thispagestyle{empty}
$~$\newpage

\superpart{Exotics}
\newpage
\thispagestyle{empty}
$~$\newpage

\input{5-cms_exo_16_022.tex}

\AddToContent{J.~Chang}
\renewcommand{\thesection}{\arabic{section}}
\newpage
\thispagestyle{empty}
$~$\newpage

\superpart{Supersymmetry}
\newpage
\thispagestyle{empty}
$~$\newpage

\input{4-cms_sus_16_041.tex}

\AddToContent{G.~Chalons, B.~Fuks, K.~Lee, J.~Park}
\renewcommand{\thesection}{\arabic{section}}

\input{6-cms_sus_17_001.tex}
\AddToContent{S.~Bein, S.-M.~Choi, B.~Fuks, S.~Jeong, D.-W.~Kang, J.~Li, J.~Sonneveld}
\renewcommand{\thesection}{\arabic{section}}
\newpage
\thispagestyle{empty}
$~$\newpage

\clearpage

\bibliography{0-biblio}

\end{document}

%% file: 0-authors.tex
\begin{center}
\textbf{Benjamin~Fuks}$^{1,2}$
\textbf{(editor)},\\
\textbf{Samuel~Bein}$^{3}$,
\textbf{Guillaume~Chalons}$^{4}$,
\textbf{Eric~Conte}$^{5}$,
\textbf{Taejeong~Kim}$^{6}$,
\textbf{Seung~J.~Lee}$^{7}$,
\textbf{Dipan~Sengupta}$^{8}$,
\textbf{Jory~Sonneveld}$^{3}$
\textbf{(convenors)},\\
Seohyun~Ahn$^{6}$,
Seungwon Baek$^{9}$,
Jung~Chang$^{10}$,
Soo-Min Choi$^{11}$,
Sihyun Jeon$^{12}$,
Sumin~Jeong$^{6}$,
Tae Hyun Jung$^{13}$,
Dong-Woo Kang$^{14}$,
Yoojin Kang$^{11}$,
Gyunggoo Lee$^{15}$,
Kyeongpil Lee$^{12}$,
Jinmian Li$^{16}$,
Jiwon~Park$^{6}$,
Jubin~Park$^{10}$,
Chaehyun~Yu$^{7}$,
Wenxing~Zhang$^{17}$,
Maxime~Zumbihl$^{1}$

\end{center}

%% file: 0-addresses.tex
{\footnotesize

$^{1}$  Sorbonne Universit\'e, CNRS, Laboratoire de Physique Th\'eorique et
        Hautes \'Energies, LPTHE, F-75005 Paris, France\\
$^{2}$  Institut Universitaire de France, 103 boulevard Saint-Michel,
        75005 Paris, France\\
$^{3}$  Institut f\"{u}r Experimentalphysik, Universit\"{a}t Hamburg,
        Luruper Chaussee 149, 22761 Hamburg, Germany\\
$^{4}$  Laboratoire de Physique Subatomique et de Cosmologie,
        Universit\'e Grenoble-Alpes, CNRS/IN2P3,\\
        53 Avenue des Martyrs, 38026 Grenoble, France\\
$^{5}$  Institut Pluridisciplinaire Hubert Curien/D\'epartement Recherches
        Subatomiques, Universit\'e de Strasbourg/CNRS-IN2P3, 23 Rue du Loess,
        F-67037 Strasbourg, France\\
$^{6}$  Department of Physics, Hanyang University, Seoul 133-791, Korea\\
$^{7}$  Department of Physics, Korea University, Seoul 136-713, Korea\\
$^{8}$  Department of Physics and Astronomy Michigan State University,
        East Lansing, MI, U.S.A.\\
$^{9}$  School of Physics, Korea Institute for Advanced Study, Seoul 130-722,
        Korea\\
$^{10}$ Department of Physics, Chonnam National University, 300 Yongbong-dong,
        Buk-gu, Gwangju, 500-757, Republic of Korea\\
$^{11}$ Department of Physics, Chung-Ang University, Seoul 06974, Korea\\
$^{12}$ Department of Physics and Astronomy, Seoul National University,
        Seoul 08826, Korea\\
$^{13}$ Center for Theoretical Physics of the Universe, Institute for
        Basic Science (IBS), Daejeon 34051, Korea\\
$^{14}$ Department of Physics \& IPAP, Yonsei University, Seoul 03722 Korea\\
$^{15}$ Department of Physics, Sungkyunkwan University, Suwon 440-746 Korea\\
$^{16}$ School of Physics, Korea Institute for Advanced Study, Seoul 130-722,
        Korea\\
$^{17}$ CAS Key Laboratory of Theoretical Physics, Institute of Theoretical
        Physics, Chinese Academy of Sciences, Beijing 100190, P. R. China\\
}

%% file: 1-intro.tex
\chapter{Introduction}
{\it S.~Bein, G.~Chalons, E.~Conte, B.~Fuks, T.~Kim, S.J.~Lee, D.~Sengupta and
  J.~Sonneveld}

The first \ma\ worskhop on LHC recasting has been held at High~1, in the Gangwon
province in South Korea on 20--27 August 2017. The workshop has brought together
a very enthusiastic group of students, postdoctoral fellows, junior as well as
more senior researchers, all interested in the development of public high-energy
physics tools allowing for the reinterpretation of the LHC results in generic
particle physics theoretical contexts. Along with the main theme of the workshop
({\it i.e.}, the problematics of the reinterpretation of the LHC searches for
new physics), various specialized lectures on collider
physics, statistics, dark matter and more formal aspects of beyond the standard
model theories have been offered, together with dedicated hands-on tutorial
sessions on the {\sc Madgraph5}~\cite{Alwall:2014hca}, {\sc Delphes}~\cite{%
deFavereau:2013fsa} and \ma~\cite{Conte:2012fm,Conte:2014zja,Dumont:2014tja}
packages.

\ma\ is a high-energy physics program that can among others be used
for the reinterpretation of the results of the LHC. It relies on an approximate
simulation of the effects of the LHC detectors through the {\sc Delphes}
framework and allows for the derivation of the number of events populating the
different signal regions of all analyses that have been implemented in its data
format. It in particular consists in a completely open source
initiative where each reimplemented analysis can be independently assigned a
Digital Object Identifier via a submission to {\sc InSpire}, ensuring that it is
uniquely identifiable, searchable and citable.

The main scope of the workshop is based on a recasting exercise assigned
to the participants. The intial group of students and postdoctoral researchers
has been divided into several subgroups of four or five people, and each
subgroup has received the task to implement, in the \ma\ framework,
a particular ATLAS or CMS search for
new physics. On top of the reimplementation task, each subgroup has been
required to assess the quality of the reimplementation through a thorough
validation procedure.
By the end of the workshop, almost all subgroups have managed to get a first
version of a \ma\ analysis code mimicking the corresponding experimental search,
along with some basic
validation of the work. For some analyses, the lack of technical information
from the experimental side has yielded slower progress, but answers to our
questions have almost always been given by the experimental groups. During
the months following the workshop, the participants have continued their work
enthusiastically, and most of the analyses have been validated and merged with
the version 1.6 of \ma.

This document summarizes the activities of the workshop and addresses in
particular the implementation and the validation, in the \ma\ framework, of
eight new ATLAS and CMS searches for new physics. If relevant, issues that have
been met are discussed, together with their impact on the quality of the
validation. The corresponding codes have
been submitted to {\sc InSpire} and are publicly available both directly within
\ma\ and from the \ma\ Public Analysis Database,\\
  \hspace*{0.7cm}%
  \url{http://madanalysis.irmp.ucl.ac.be/wiki/PublicAnalysisDatabase}.

This document is divided into three parts according to the classes of analyses
under consideration. In the first of these parts, one focuses on LHC
searches for dark matter in varied channels. We consider two searches for a
mono-Higgs signal, one from ATLAS~\cite{Aaboud:2017yqz} and one from CMS~\cite{%
Sirunyan:2017hnk}, in which a Higgs boson is assumed to be produced with a pair
of dark matter particles manifesting themselves as missing energy in the
detector. We moreover recast one ATLAS search dedicated to the production of a
hard photon in association with missing energy~\cite{Aaboud:2017dor}, one ATLAS
search for
dark matter production in association with light jets~\cite{Aaboud:2017phn}
and heavy-flavor jets~\cite{ATLAS:2016tsc}. In the second part of this document,
we detail a more exotic CMS search for long-lived electrons and muons~\cite{%
CMS:2016isf}, which has required the development of new features within
\ma. Finally, in the last part of these proceedings, we detail more classical
searches for supersymmetric particles, first in the multilepton plus jets plus
missing transverse energy channel~\cite{Sirunyan:2017hvp}, and next in the
opposite-sign same-flavor dilepton case~\cite{Sirunyan:2017leh}.

%% file: 8-atlas_conf_2016_086.tex
\chapter{ATLAS-CONF-2016-086: an ATLAS dark matter search with $b$-jets and
  missing energy (13.3~fb$^{-1}$)}
\label{atlas_conf_2016_086}
{\it B.~Fuks, M.~Zumbihl}

\begin{abstract}
We present the \ma\ implementation and validation of the ATLAS-CONF-2016-086
search. This ATLAS analysis targets the production of dark matter in association
with $b$-tagged jets and probes 13.3~fb$^{-1}$ of LHC proton-proton collisions
at a center-of-mass energy of 13~TeV. The validation of our reimplementation is
based on a comparison with all the material provided by the ATLAS collaboration,
as well as with a back-of-the-enveloppe expectation of a related theoretical
work. By lack of public experimental information, we have not been able to
validate this reimplementation more throroughly.
\end{abstract}

\section{Introduction}

In this note, we describe the validation of the implementation, in the \ma\
framework~\cite{Conte:2012fm,Conte:2014zja,Dumont:2014tja}, of the
ATLAS-CONF-2016-086 analysis~\cite{ATLAS:2016tsc} probing the production of dark
matter at the LHC in
association with a pair of $b$-tagged jets originating from a bottom-antibottom
quark pair at the parton level. The signature that is searched for thus consists
in missing transverse energy and $b$-jets. The ATLAS-CONF-2016-086 analysis
focuses on the analysis of an integrated luminosity of 13.3~fb$^{-1}$ of LHC
collisions at a center-of-mass energy of 13~TeV.

For the validation of our reimplementation, we have focused on a
simplified dark matter model in which the Standard Model is extended by two
additional fields, namely a Dirac field $\chi$ corresponding to the dark matter
particle and a scalar ($\Phi$) or pseudoscalar ($A$) field responsible for the
mediation of the interactions of the Standard Model sector with the dark
sector~\cite{Abercrombie:2015wmb}. This scenario involves four parameters,
namely the mass of the scalar mediator
$m_{\Phi}$ (or $m_A$ in the pseudoscalar case), the mass of the dark matter
particle $m_{\chi}$, the mediator coupling to the dark sector $y_\chi$ and the
flavor-universal coupling of the mediator to the Standard Model $y_v$. In this
theoretical framework, the signal that is relevant for the considered analysis
arises from the process
\be
  pp \rightarrow \chi\bar{\chi}\ b\bar{b} \ ,
\ee
in which the pair of dark matter particles gives rise to missing transverse
energy and originates from the decay of a possibly off-shell mediator.

\section{Description of the analysis}

The analysis makes use of all the information present in the signal final state.
It therefore requires, as a basic selection, the presence of missing
transverse energy as well as of jets with some of them being $b$-tagged. The
kinematics of the bottom-antibottom system is then used as a handle to reduce
the background of the Standard Model.

\subsection{Object definitions}

Jets are recontructed by means of the anti-$k_T$ algorithm~\cite{%
Cacciari:2008gp} with a radius parameter set to $R=0.4$. Our analysis focuses on
jets whose transverse momentum $p_T^j$ and pseudorapidity $\eta^j$ fullfill
\be
  p_T^j>20~{\rm GeV} \qquad\text{and}\qquad |\eta^j|<2.8 \ .
\ee
Moreover, the selected jets are tagged as originating from the fragmentation of
a $b$-quark according to a working point for which the average $b$-tagging
efficiency is of about 60\%.

Electron candidates are required to have a transverse momentum $p_T^e$ and
pseudorapidity $\eta^e$ obeying to
\be
  p_T^e >7~{\rm GeV} \qquad\text{and}\qquad |\eta^e|<2.47\ ,
\ee
and the muon candidate definition is similar, although with slightly looser
thresholds,
\be
  p_T^\mu >6~{\rm GeV} \qquad\text{and}\qquad |\eta^\mu|<2.7\ .
\ee
Any jet lying within a cone of radius $\Delta R < 0.2$ of an electron is
discarded, unless it
is $b$-tagged. In this last case, it is the electron that is discarded. Any
electron or muon that would then lie within a cone of radius $\Delta R < 0.4$ of
a jet is finally removed from the set of jet candidates to consider.

The missing transverse momentum vector $\slashed{\bf p}_T$ is defined as the
opposite of the vector sum of the momenta of all reconstructed physics object
candidates, and the missing transverse energy $\slashed{E}_T$ is then defined by
its norm.

\subsection{Event selection}
The analysis contains a unique signal region that is defined by a requirement on
the missing transverse energy,
\be
  \slashed{E}_T > 150~{\rm GeV} \ ,
\ee
and on the number of $b$-tagged jets that is asked to be equal to 2. The signal
being
charaterized by a small jet multiplicity, events featuring a third jet with a
transverse momentum greater than 60~GeV are vetoed, as events whose final
state contains leptons. Moreover, the missing transverse momentum is constrained
to be well separated from any jet,
\be
  \Delta\phi(\slashed{\bf p}_T, j) > 0.4 \ .
\ee

In order to guarantee a full trigger efficiency, selected events are
required to satisfy the so-called hyperbolic requirement on the missing energy,
\be
  p_T^{j_1} > 85~{\rm GeV} \qquad\text{and}\qquad
  \slashed{E}_T > \frac{\big[150~{\rm GeV}\big] p_T^{j_1} -
      \big[11700~{\rm GeV}^2\big]}
      {p_T^{j_1} - \big[85~{\rm GeV}\big]} \ .
\ee
The dominant component of the background, related to invisible $Z$-boson
production in association with $b$-tagged jets, is reduced by requiring a large
separation between the jet candidates,
\be
  \Delta R(j_i,j_k) > 2.8,
\ee
for any pair of reconstructed jets $(i,k)$. The two $b$-jets are furthermore
constrained to satisfy
\be
  \Delta\eta(b_1,b_2) > 0.5\ , \qquad
  \Delta\phi(b_1,b_2) > 2.2 \qquad\text{and}\qquad
  {\rm Imb}(b_1,b_2) \equiv
    \frac{p_T^{b_1} - p_T^{b_2}}{p_T^{b_1} + p_T^{b_2}} > 0.5 \ .
\ee
With the last requirement, one imposes a significant transverse-momentum
imbalance between the two $b$-jets that is known to be large for typical
signals.

\section{Validation}
\subsection{Event Generation}
In order to validate our analysis, we rely on the dark matter simplified model
introduced above and for which a UFO model~\cite{Degrande:2011ua} has been
provided by the ATLAS collaboration. We focus on a benchmark scenario defined by
\be
  y_{\chi} = y_v = 1,  \qquad
  m_{\Phi/A} = 20~{\rm GeV} \qquad\text{and}\qquad m_{\chi} = 1~{\rm GeV}.
\ee
We make use of {\sc MadGraph5\_aMC@NLO} version~2.6.0~\cite{Alwall:2014hca} for
hard-scattering event generation in which leading-order matrix elements are
convoluted with the leading-order set of NNPDF~3.0 parton
densities~\cite{Ball:2014uwa}. Those events have been showered by means of the
{\sc Pythia}~6 package~\cite{Sjostrand:2006za}. Finally, the simulation of the
detector response has been performed by using
{\sc Delphes}~3~\cite{deFavereau:2013fsa}, that relies on
{\sc FastJet}~\cite{Cacciari:2011ma} for object reconstruction and that has been
used with an appropriate tuned detector card. All
necessary configuration files, most of them having been provided by ATLAS, can
be found on the \ma\ public database webpage,\\
\hspace*{0.5cm}\url{http://madanalysis.irmp.ucl.ac.be/wiki/PublicAnalysisDatabase}.\\
We have finally used our \ma\ reimplementation to calculate the signal selection
efficiencies.

\subsection{Comparison with the official results}
\begin{figure}
    \centering
    \includegraphics[width=0.48\columnwidth]{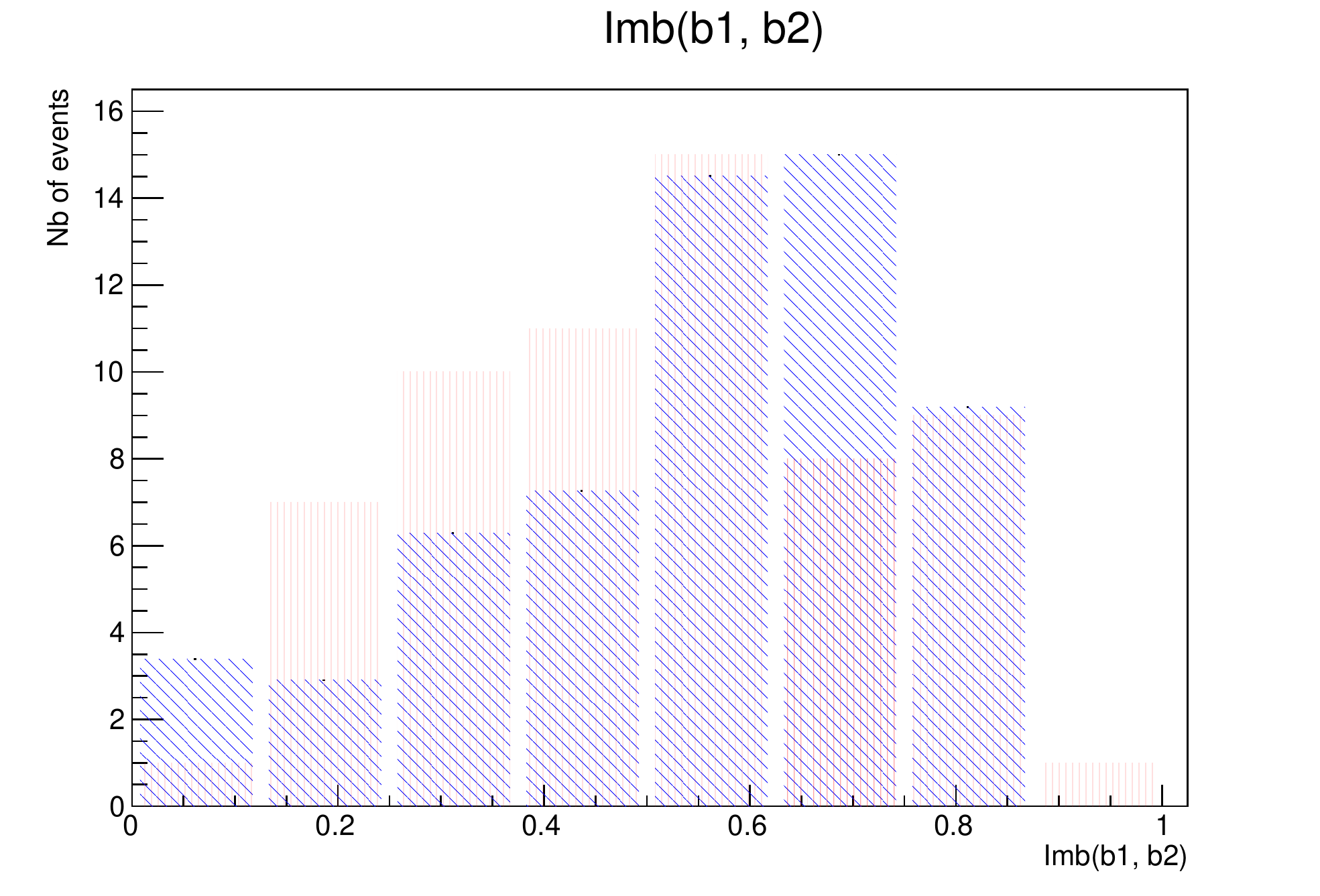}
    \includegraphics[width=0.48\columnwidth]{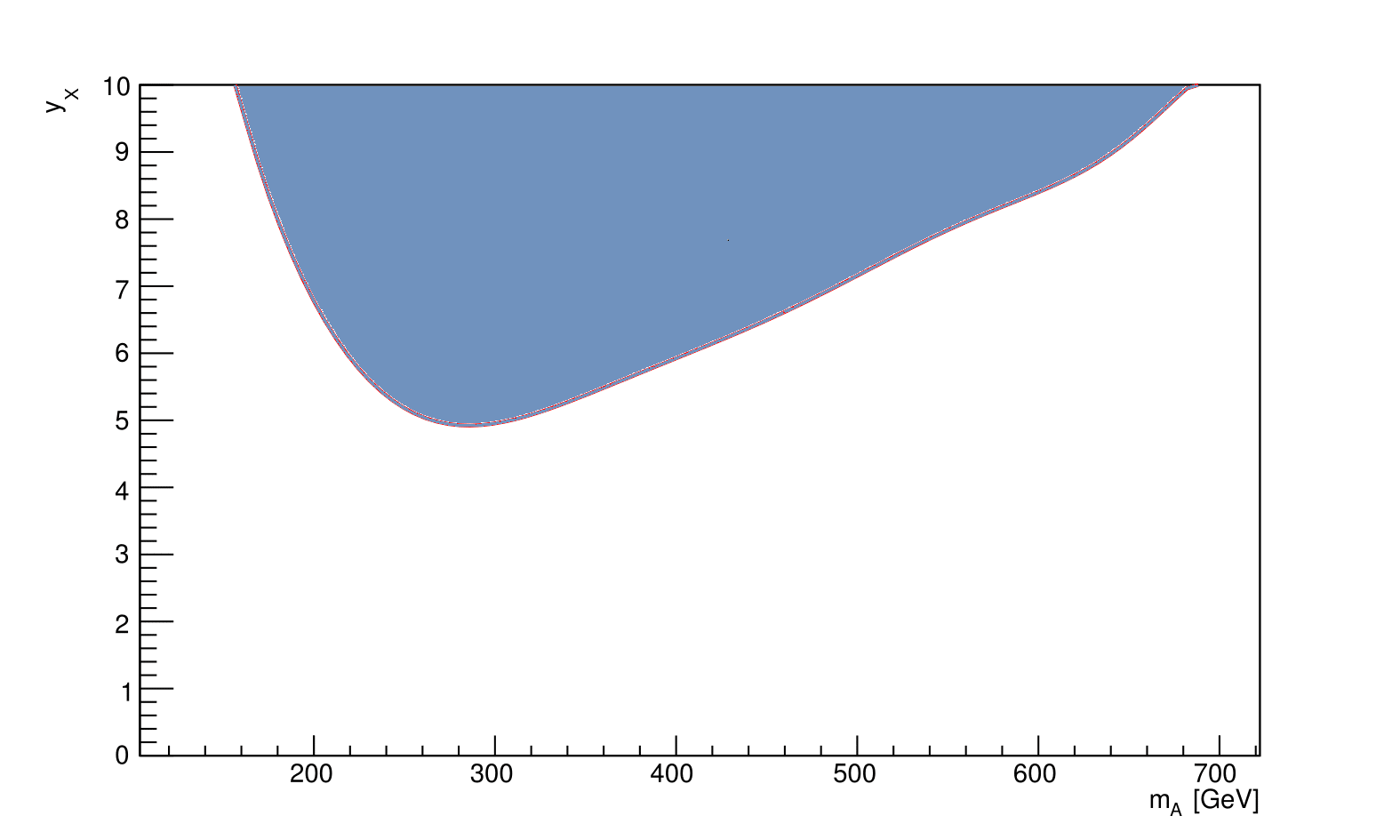}
    \caption{Left: Transverse momentum imbalance when all the analysis selection
      criteria are applied, except the one on Imb$(b_1, b_2)$. We compare the
      official numbers (red) with our predictions (blue). Right: Region of the
      parameter space of the new physics model introduced in Ref.~\cite{%
      Banerjee:2017wxi} excluded at the 95\% confidence level for
      new physics scenarios in which $m_\chi=100$~GeV and $y_v=1$.
    \label{fig:imb}}
\end{figure}

In the left panel of
Figure~\ref{fig:imb}, we present the transverse-momentum imbalance spectrum
as computed using the \ma\ (blue) and compare it to the official results (red).
The results shown in the figure include all selection cuts but the Imb$(b_!,
b_2)$ one. One obtains a fair agreement accounting for the large statistical
uncertainties plaguing the simulation and about which no information has been
provided by the ATLAS collaboration.

In the right panel of the figure, we consider a different new physics setup in
which the dark matter mass is set to $m_\chi = 100$~GeV and the mediator
coupling to the Standard Model to $g_v=1$. We then present, in the
$(m_A, y_\chi)$ plane, the parameter space region that is excluded at the 95\%
confidence level. We obtain a good agreement with the back-to-the-enveloppe
estimations of Ref.~\cite{Banerjee:2017wxi}.

\section{Summary}
We have implemented the ATLAS-CONF-2016-086 search in the \ma\ framework. Our
analysis has been validated in the context of a simpified model for dark matter
in which the dark matter candidate is a fermion and the mediator a boson. We
have found a decent agreement with the material provided by ATLAS, which is
not dramatically detailed. Due to the lack of information, the validation has
been kept brief. As a fair agreement has nevertheless been obtained both with
respect to the material provided by ATLAS and to an earlier theoretical work, we
have considered this reimplementation as validated. It is available from \ma\
version 1.6 onwards, its Public Analysis Database and from {\sc InSpire}~\cite{%
1635567},\\
\hspace*{0.7cm}\url{http://doi.org/10.7484/INSPIREHEP.DATA.UUIF.89NC}.

%% file: 7-atlas_1707_01302.tex
\chapter{ATLAS-EXOT-2016-25: an ATLAS mono-Higgs analysis (36.1~fb$^{-1}$)}
\label{atlas_1707_01302}
{\it S.~Jeon, Y.~Kang, G.~Lee, C.~Yu}

\begin{abstract}
We present the \ma\ implementation and validation of the ATLAS-EXOT-2016-025
analysis, which concerns a search for dark matter when it is produced in
association with a Higgs boson decaying into a $b\bar b$ system. The results
consider a dataset of proton-proton collisions at a center-of-mass energy of
13~TeV corresponding to an integrated luminosity of 36.1~fb$^{-1}$, as
recorded by the ATLAS collaboration during the LHC Run~2. The validation of our
reimplementation is based on a comparison of our predictions with official ATLAS
numbers in the context of a new physics scenario featuring two Higgs doublets,
an extra gauge boson and a dark matter particle. A good agreement has been
found for the light new physics case, but issues have occurred for heavier new
particles. The ATLAS collaboration has not provided any information allowing us
to understand the problems deeper.
\end{abstract}

\section{Introduction}
In this note, we describe the validation of our implementation of an ATLAS dark
matter search in the \ma\ framework~\cite{Conte:2012fm,Conte:2014zja,%
Dumont:2014tja}. This analysis, dubbed ATLAS-EXOT-2016-25, performs a search for
dark matter production in association with a Higgs boson ($h$) decaying into a
pair of $b$ quarks~\cite{Aaboud:2017yqz}. It relies on 36.1 fb$^{-1}$ of data
recorded by the ATLAS detector from LHC proton-proton collisions at a
center-of-mass energy of 13~TeV. The search focuses on two regimes, respectively
targetting a resolved Higgs boson where its decay products can be distinguished
and a merged regime in which the Higgs boson decays into a single fat jet. We
focus here only on the resolved regime due to a lack of experimental
information on the merged regime.

Our validation relies on a reinterpretation of the ATLAS results of the analysis
in a dark matter $Z'$-Two-Higgs-Doublet model in which the Standard Model is
supplemented by a dark matter particle $\chi$, a $Z'$ boson and a second
Higgs doublet~\cite{Berlin:2014cfa,Abercrombie:2015wmb}. The signal under
consideration corresponds to the resonant production of a $Z'$ boson that then
decays into a Standard Model Higgs boson $h$ and a pseudoscalar boson $A^0$. The
latter play the role of a portal to the dark sector, and thus decays invisibly
into two dark matter particles. The process under consideration hence reads
\be
  p p \to Z' \to h A^0 \to h \chi\chi \ .
\ee

\section{Description of the analysis}
This analysis selection is strictly based on the considered signature and
requires the presence of a significant amount of missing transverse energy
(carried by the dark matter particle), well separated from the jet activity
associated with the Higgs boson. The analysis moreover asks for at least two
hard jets that are compatible with the decay of the Higgs boson, with at least
one of them being $b$-tagged.

\subsection{Object definitions and preselection}
The analysis mainly relies on jets, that are reconstructed following the
anti-$k_T$ algorithm~\cite{Cacciari:2008gp}, with a radius parameter set to
$R=0.4$. Jets with a transverse momentum $p_T^j$ and pseudorapidity $\eta^j$
satisfying
\be
  p_T^j > 20~{\rm GeV} \qquad\text{and}\qquad
  |\eta^j| < 2.5
\ee
are denoted as central jets and those for which
\be
  p_T^j > 30~{\rm GeV} \qquad\text{and}\qquad
  2.5 < |\eta^j| < 4.5
\ee
are called forward jets. Whilst the analysis also makes use of jets
reconstructed with the anti-$k_T$ algorithm~\cite{Cacciari:2008gp} and a radius
parameter fixed to $R=1$, these are connected to the merged regime where the
Higgs boson is boosted and that we were not able to validate by virtue of the
lack of ATLAS information. We have thus ignored them.
Electron candidates are required to have a transverse momentum $p_T^e$ and
pseudorapidity $\eta^e$ obeying to
\be
  p_T^e >7~{\rm GeV} \qquad\text{and}\qquad |\eta^e|<2.47\ ,
\ee
while muon candidates are similarly defined, although the thresholds are
slightly looser,
\be
  p_T^\mu >7~{\rm GeV} \qquad\text{and}\qquad |\eta^\mu|<2.7\ .
\ee
In both cases, loose isolation criteria have been imposed~\cite{Aad:2016jkr,%
Aaboud:2016obm}. Moreover, any jet lying at an angular distance in the
transverse plane $\Delta R\leq 0.2$ of an electron has
been removed.

The missing transverse momentum vector ${\bf E}_T^{\rm miss}$ is defined as the
opposite of the vector sum of the momenta of all reconstructed physics object
candidates, and the missing transverse energy is defined by its norm
\be
  E_T^{\rm miss}=|{\bf E}_T^{\rm miss}|\ .
\ee

\subsection{Event Selection}

We focus on the resolved Higgs regime for which a single signal region is
defined. It requires
\be
  150~{\rm GeV} < E_T^{\rm miss} < 500~{\rm GeV,}
\ee
a criterion that also allows the missing-energy-only trigger to be
fully efficient. In order to suppress the multijet background, the missing
transverse momentum is constrained to be well separated in azimuth from the
three leading jets (if relevant),
\be
   \Delta \phi({\bf E}_T^{\rm miss}, {\bf p}_T^j) > \frac{\pi}{9} \ ,
\ee
and more or less aligned with the missing transverse momentum recontructed from
the tracker information only ${\bf p}_T^{\rm miss, trk}$,
\be
 \Delta \phi({\bf E}_T^{\rm miss}, {\bf p}_T^{\rm miss,trk}) < \frac{\pi}{2} \ .
\ee
In addition, this last quantity is required to fullfil
\be
  \big| {\bf p}_T^{\rm miss,trk} \big| > 30~{\rm GeV}.
\ee
The analysis requires the presence of at least two jets,
\be
  N_j > 2 \ ,
\ee
with either one or two of them being $b$-tagged, and at least one of
them featuring a transverse
momentum larger than 45~GeV,
\be
  p_T^{j_1} > 45~{\rm GeV.}
\ee
We have restricted our reimplementation procedure to the case
\be
  N_b = 2 \ ,
\ee
as this region is expected to be the most sensitive to the signal. It
additionally consists of the only signal region for which validation material
has been provided. These two
$b$-jets are then considered as the Higgs system. As the Higgs system lies in a
configuration in which it is recoiling
against a pair of dark matter particle, one requires
\be
  \Delta \phi({\bf E}_T^{\rm miss}, {\bf p_T}_h) > \frac{2 \pi}{3} \ ,
\ee
where ${\bf p}_T^h$ denotes the transverse momentum of the reconstructed Higgs
boson. Moreover, the scalar sum of the transverse momentum of the two and three
leading jets ($H_{T,2j}$ and $H_{T,3j}$) is imposed to satisfy
\be
  H_{T,2j} > 120~{\rm GeV}
  \qquad\text{and}\qquad
  H_{T,3j} > 150~{\rm GeV} \ ,
\ee
this last requirement being imposed only if at least three central jets are
present.

In order to optimize the selection, the two jets $j_1$ and $j_2$
defining the Higgs system are enforced to be not too separated,
\be
  \Delta\phi(j_1,j_2) < \frac{7\pi}{9}
  \qquad\text{and}\qquad
  \Delta R(j_1,j_2) < 1.8 \ ,
\ee
and a tau lepton veto is imposed. As an additional selection, the scalar sum of
the transverse momentum of the $j_1$ and $j_2$ jets, as well as of the
third jet if present, is required to satisfy
\be
  p_T^{j_1} + p_T^{j_2} (+ p_T^{j_3}) < 0.63 H_T \ ,
\ee
where the hadronic activity $H_T$ in the event consists in the scalar sum of the
transverse momentum of all reconstructed jets.

\section{Validation}

\subsection{Event generation}
In order to validate our reimplementation, we consider two benchmark scenarios
in which the $Z'$-boson mass $m_{Z'}$ is respectively fixed to 600~GeV and
1400~GeV. Correspondingly, the pseudoscalar mass $m_{A^0}$ is fixed to 300~GeV
and 600~GeV. In all
cases, the mass of the dark matter particle is taken vanishing.

We have made use of \textsc{MadGraph5}\_aMC@NLO~\cite{Alwall:2014hca} for
generating hard-scattering signal events, relying on the
UFO~\cite{Degrande:2011ua} model shared by the ATLAS collaboration. The
generated matrix element has been convoluted with the next-to-leading-order set
of NNPDF~3.0 parton densities~\cite{Ball:2014uwa}, and we have handled the Higgs
into $b\bar b$ decay,
parton showering and hadronization with {\sc Pythia}~8\cite{Sjostrand:2014zea}.
The simulation of the response of the ATLAS detector is achieved via
\textsc{Delphes 3}~\cite{deFavereau:2013fsa}, that
internally relies on {\sc FastJet}~\cite{Cacciari:2011ma} for object
reconstruction, with an tuned detector configuration.

\subsection{Comparison with the official results}
\begin{figure}
  \centering
  \includegraphics[width=0.65\columnwidth]{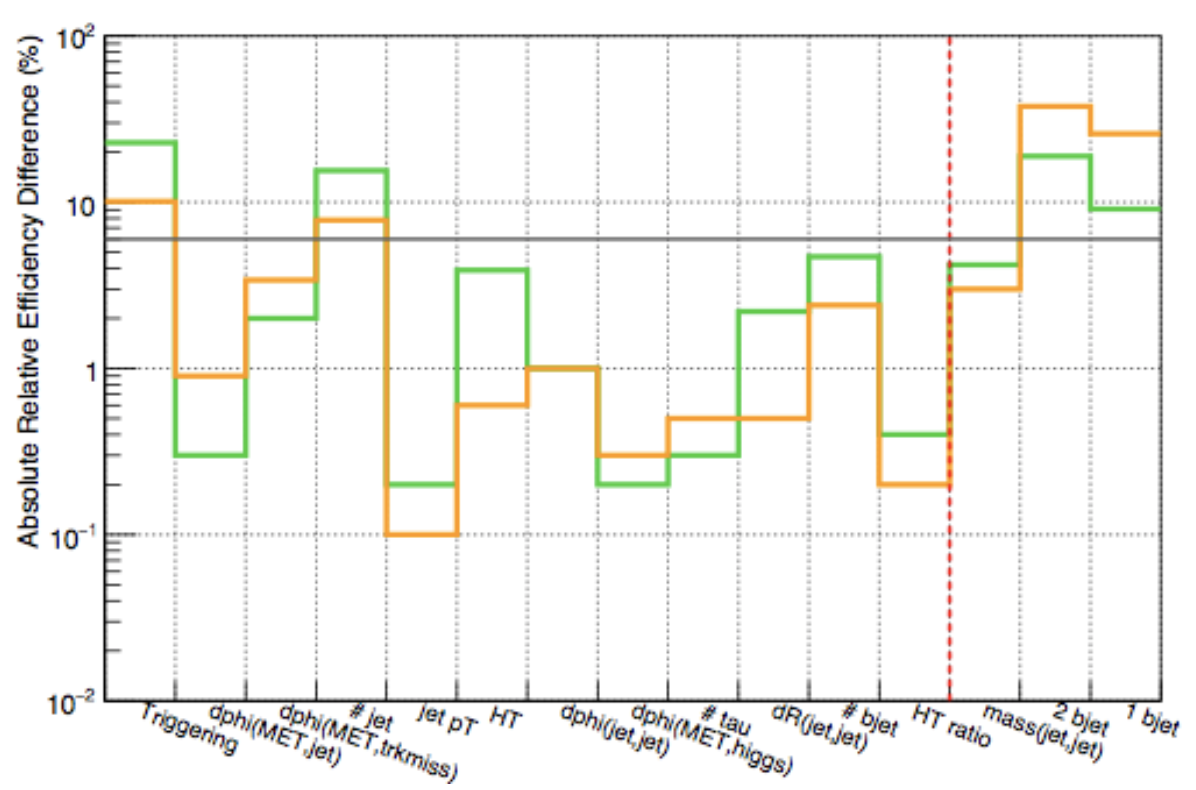}
  \caption{Relative difference between the ATLAS official and \ma\ predictions
    for the efficiency of each selection cut, for two benchmarks
    defined by $(m_{Z'}, m_{A^0}) = (600, 300)$~GeV (green) and $(1400, 600)$~%
    GeV (orange). The solid horizontal line indicates a 6\% difference reference
    line.}
  \label{fig:7-efficiencies}
\end{figure}

In Figure~\ref{fig:7-efficiencies}, we present the relative difference between
the \ma\ predictions and the ATLAS official results for the two considered
scenarios, computed as
\be
 \delta =  1-\frac{\epsilon_i^{\rm MA5}}{\epsilon_i^{\rm ATLAS}} \ ,
\ee
where the index $i$ corresponds to the cut number, and where
$\epsilon^{\rm MA5}_i$ and $\epsilon^{\rm ATLAS}_i$ indicate the predicted and
ATLAS efficiencies for the cut number $i$. The results include two extra
cuts, available in the validation material. The Higgs system invariant
mass is firstly imposed to satisfy
\be
  50~{\rm GeV} < m_{j_1 j_2} < 250~{\rm GeV} \ ,
\ee
so that it is loosely compatible with a Higgs boson, and one secondly imposes
either one
or two $b$-tag requirements. For what concerns the last three cuts, only one of
them is imposed at a time.

The large differences at the level of the trigger (first cut) is expected, as
not all requirements, and in particular the features at the level of the turn-on
of the trigger efficiency curve near threshold, can be implemented in
{\sc Delphes}. Moreover, large discrepancies are also observed for the last
selections that strongly rely on jets. After discussions with ATLAS, it turned
out that our reimplementation were not matching well what ATLAS actually
implemented. However, the corresponding information was lost (within ATLAS) and
we have never been able to understand the origins of the differences.

\begin{table}
\centering
 \renewcommand{\arraystretch}{1.4}
\begin{tabular}{l||l l l| l l l}
   \multirow{2}{*}{Cuts} & \multicolumn{3}{c|}{$(m_{Z'}, m_{A^0}) =
     (600, 300)$~GeV} & \multicolumn{3}{c}{$(m_{Z'}, m_{A^0}) = (1400,
     600)$~GeV}\\
                              & MA5     & Official & error &
   MA5     & Official & error \\ \hline
  $E_T^{\rm miss}$           & 0.772   & 0.89     & 13.3\% &
    0.660 & 0.604 & 9.2\%\\
  ${\bf p}_T^{\rm miss,trk}$ & 0.757   & 0.711    &  6.5\% &
    0.657 & 0.546 & 20.3\%\\
  $\Delta \phi({\bf E}_T^{\rm miss}, {\bf p}_T^j)$ & 0.727 & 0.685 & 6.1\% &
    0.592 & 0.497 & 19.1\%\\
  $\Delta\phi({\bf E}_T^{\rm miss}, {\bf p}_T^{\rm miss,trk})$ & 0.727 & 0.671
    & 8.3\% & 0.592 & 0.480 & 23.3\%\\
  $N_j$   & 0.602 & 0.658 & 8.5\% & 0.523 & 0.460 & 13.7\%\\
  $p_T^j$ & 0.599 & 0.655 & 8.5\% & 0.522 & 0.459 & 13.7\%\\
  $H_T$   & 0.572 & 0.651 & 12.1\% & 0.519 & 0.459 & 13.1\%\\
  $\Delta\phi(j_1,j_2)$ & 0.556 & 0.633 & 12.2\% & 0.494 & 0.441 & 12.0\%\\
  $\Delta\phi({\bf E}_T^{\rm miss}, {\bf p_T}_h)$ & 0.544 & 0.620 & 12.3\%&
    0.490 & 0.439 & 11.6\%\\
  tau veto & 0.530 & 0.603 & 12.1\% & 0.476 & 0.424 & 12.3\%\\
  $\Delta R(j_1,j_2)$ & 0.455 & 0.506 & 10.0\% & 0.434 & 0.385 & 12.7\%\\
  $1\leq N_b \leq 2$ & 0.431 & 0.503 & 14.1\% & 0.421 & 0.383 & 9.9\%\\
  $\sum p_T^j$ & 0.430 & 0.499 & 13.8\% & 0.421 & 0.382 & 10.2\%\\
  \hline
  $m_{j_1 j_2}$ & 0.396 & 0.481 & 17.7\% & 0.404 & 0.376 & 7.4\%\\
  2 $b$-jets & 0.252 & 0.246 & 2.4\% & 0.269 & 0.177 & 52.0\%\\
  1 $b$-jet & 0.154 & 0.197 & 21.8\% & 0.135 & 0.165 & 18.2\%
\end{tabular}
\caption{Comparison of the cutflow predicted by \ma\ with the one provided by
  the ATLAS collaboration for the $(m_{Z'}, m_{A^0}) = (600, 300)$~GeV benchmark
  scenario (left) and $(m_{Z'}, m_{A^0}) = (1400, 600)$~GeV benchmark scenario
  (right).}
\label{tab:cutflows}
\end{table}

In general, our reimplementation nevertheless performs quite well, in particular
in terms of the total selection efficiencies and for benchmark scenarios
featuring light particles. This is illustrated in Table~\ref{tab:cutflows}
(left), where we present the total selection efficiencies on a
cut-by-cut basis. For the $(m_{Z'}, m_{A^0}) = (600, 300)$~GeV scenario, we
observe that an agreement of order of 10-20\% all along
the selection (left part of the table). However, for heavier scenarios, we
have found larger discrepancies. The ATLAS collaboration has however not been
able to provide information allowing us to understand these discrepancies,
except that our {\sc Delphes} tuning may be incorrect in the large $p_T$ range.
The collaboration has however not provided any additional information allowing
us to fix the issue.

We remind that the `1 $b$-jet' and `$m_{j_1 j_2}$' validation regions
have not been implemented into our the code, as they correspond to additional
cuts that have been implemented solely for validation purposes. The signal
region of interest focuses instead on the `$N_b=2$' case.

\section{Conclusion}
We have implemented in \ma\ a mono-Higgs analysis performed by the ATLAS 
collaboration and have tried to validate our implementation in the context of a
Two-Higgs-Doublet model featuring an extra neutral gauge boson and a dark
matter particle. After having compared our results with the official ones, we
have found that our reimplementation was trustable for light new physics
scenarios, but not for heavier cases. We therefore recommend caution when using
this analysis for phenomenological purposes. As a fair agreement has been
obtained in the light case, so that our reimplemented analysis could be used for
such scenarios, we have considered this reimplementation (partly) validated and
have made it available from \ma\ version 1.6 onwards and its Public Analysis
Database and from {\sc InSpire}~\cite{%
1672227},\\
\hspace*{0.7cm}\url{http://doi.org/10.7484/INSPIREHEP.DATA.SSS4.298U}.

\newpage$~$
\thispagestyle{empty}
\newpage

%% file: 9-atlas_1711_03301.tex
\chapter{ATLAS-EXOT-2016-27: an ATLAS monojet analysis (36.2 fb$^{-1}$)}
\label{atlas-1711-03301}
{\it D.~Sengupta}

\begin{abstract}
We present the \ma\ implementation of the recent ATLAS-EXOT-2016-27 monojet
search. This search allows us to probe various new physics scenarios featuring a
dark matter particle through the so-called monojet channel in which the
final-state signature consists in one highly-energetic jet recoiling against
missing transverse energy carried by dark matter particles. The results are
based on the analysis of a dataset of 36.2~fb$^{-1}$ of proton-proton collisions
recorded by the ATLAS detector with a center-of-mass energy of 13~TeV. The
validation of our reimplementation relies on a comparison of our predictions
with the official ATLAS results in the context
of a supersymmetry-inspired simplified model in which the Standard Model is
extended by a neutralino and a stop decaying into a charm
quark and a neutralino.
\end{abstract}

\section{Introduction}
In this contribution, we present the validation of the implementation, in the
\ma~\cite{Conte:2012fm,Conte:2014zja,Dumont:2014tja} framework, of the
ATLAS-EXOT-2016-27 search for dark matter in the monojet
channel~\cite{Aaboud:2017phn}. This search is in particular sensitive to certain
supersymmetric scenarios, dark matter setups and extra dimensional models. Each
of those models can indeed predict, in specific realizations, the production of
a pair of invisible particles in association with a highly-energetic jet ({\it
i.e.}, the signature under consideration).

\begin{figure}[t]
  \centering
    \includegraphics[width=0.30\textwidth]{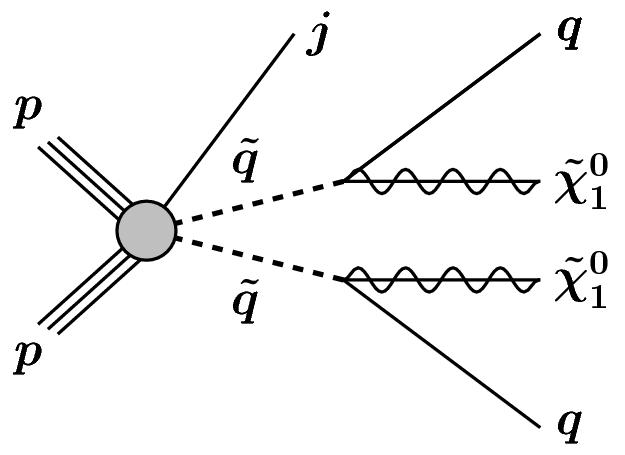}
  \caption{Representative Feynman diagram corresponding to the production of a
    pair of squarks $\tilde q$ that each decays into a neutralino
    $\tilde\chi_1^0$ and a light quark $q$.}
  \label{fig:9-diagrams}
\end{figure}

For our validation procedure, we focus on a compressed supersymmetric
configuration in which the searched for signature arises from the associated
production of a hard jet with a pair of invisible squarks that each decays into
a soft light jet and a neutralino. This process is illustrated by the
representative Feynman diagram of Fig.~\ref{fig:9-diagrams}. The considered
analysis is in particular sensitive to the case of a compressed light stop that
decays into a charm quark and a neutralino (through a flavor-violating
loop-induced subprocess),
\be
 p p \to j\ \tilde t^\ast \tilde t
     \to j\ c \tilde\chi_1^0 \bar c \tilde\chi_1^0 \ .
\ee
This decay mode of the top quark becomes especially relevant when the more
standard decay channels involving either a top quark or a chargino are closed.

\section{Description of the analysis}
\begin{figure}[t]
  \centering
    \includegraphics[width=1.00\textwidth]{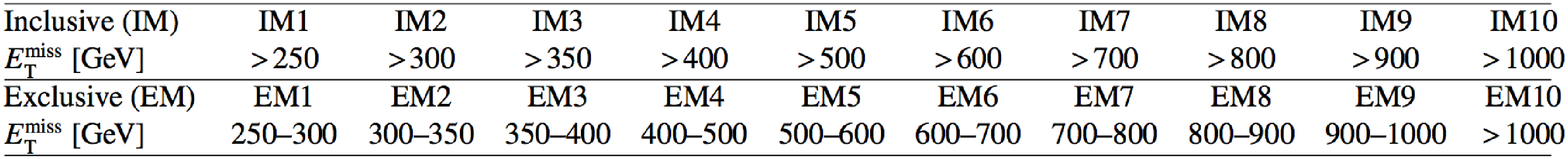}
  \caption{Missing transverse energy requirements of the 20 signal regions of
    the ATLAS-EXOT-2016-27 analysis.}
  \label{9-signalregion}
\end{figure}

The ATLAS monojet analysis targets a final-state containing at least one very
energetic jet that is assumed to originate from initial state radiation, as well
as a certain amount of missing transverse energy $E_T^{\rm miss}$. The analysis
strategy is twofold, depending on the selection cut on the missing transverse
energy. In a first series of ten signal regions (IM1, IM2, $\ldots$, IM10), it
considers inclusive missing transverse energy selections,
\be
  E_T^{\rm miss} > E_{\rm threshold}\ ,
\ee
where the 10 different thresholds range from 250~GeV to 1~TeV, as shown on the
first line of the table of Fig.~\ref{9-signalregion}. In a second series of
signal regions, the analysis instead considers exclusive missing tranverse
energy selection,
\be
  E_{\rm threshold}^{\rm min}\le E_T^{\rm miss}\le E_{\rm threshold}^{\rm max}
   \ .
\ee
The thresholds associated with the 10 corresponding signal regions (EM1,
EM2, $\ldots$, EM10) are shown in the second table of Fig.~\ref{9-signalregion}.

\subsection{Object definition}
Jets are recontructed following the anti-$k_T$ algorithm~\cite{Cacciari:2008gp}
with a radius parameter $R=0.4$, and only those jets
with a transverse momentum $p_T^j$ and pseudorapidity $\eta^j$ satisfying
\be
  p_T^j>20~{\rm GeV} \qquad\text{and}\qquad |\eta^j|<2.8
\ee
are retained. Among those jets, those with a transverse momentum greater than
30~GeV and with a pseudorapidity smaller than 2.5 (in absolute value) are
potentially considered as $b$-tagged, according to a $b$-tagging working point
that is in average 60\% efficient~\cite{ATL-PHYS-PUB-2016-012}.

Electron candidates are required to have a transverse momentum $p_T^e$ and
pseudorapidity $\eta^e$ obeying to
\be
  p_T^e >20~{\rm GeV} \qquad\text{and}\qquad |\eta^e|<2.47\ ,
\ee
whereas muon candidates must obey to
\be
  p_T^\mu > 10~{\rm GeV} \qquad\text{and}\qquad |\eta^\mu|<2.7\ .
\ee
Any non-$b$-tagged jet with $p_T^j>30$~GeV lying within a cone of radius $\Delta
R < 0.2$ from an electron is discarded, whilst any electron lying within a cone
of radius $\Delta R < 0.2$ centered on a $b$-tagged jet is removed. Any electron
that would then lie within a cone
of radius $0.2 < \Delta R < 0.4$ of a jet is finally removed in a second step.
In addition, jets with a $p_T^j > 30$~GeV are discarded if they are lying in a
cone of radius $\Delta R < 0.4$ centered on any muon.

The missing transverse momentum vector $\slashed{\bf p}_T$ is defined as the
opposite of the vector sum of the momenta of all reconstructed physics object
candidates with a pseudorapidity smaller than 4.9, and the missing transverse
energy $E_T^{\rm miss}$ is defined by its norm.

\subsection{Event Selection}

Event preselection imposes first the presence of a significant amount of missing
energy,
\be
  E^{\rm miss}_T >250~{\rm GeV},
\ee
and next that the final state features a monojet-like topology, the leading jet
being imposed to satisfy
\be
  p_{T}(j_1) > 250~{\rm GeV}.
\ee
Electron and muon vetos are then enforced, and any jet $j$ has to be well
separated from the missing momentum,
\be
  \Delta\phi(j, \slashed{\bf p}_{T}) > 0.4 \ .
\ee
Selected events are then categorized into the inclusive and exclusive signal
regions introduced in Fig.~\ref{9-signalregion}.

\section{Validation}

For our validation, we generate events for various simplified models inspired by
the MSSM. We consider a class of models where the Standard Model is extended by
a stop (of mass $M_{\tilde t}$) and a neutralino (of mass $M_{\tilde \chi}$),
all other superymmetric states being taken decoupled. For each choice of mass
parameters, our signal event samples are normalized to an
integrated luminosity of 36.2~fb$^{-1}$ and to a cross section evaluated at the
NLO+NLL accuracy~\cite{Borschensky:2014cia}.

Signal events have been generated with \textsc{MadGraph5\_aMC@NLO}~%
\cite{Alwall:2014hca} and {\sc Pythia 8}~\cite{Sjostrand:2014zea} for the hard
scattering matrix elements and the simulation of the parton showering and
hadronization, respectively. We have considered event samples describing final
states featuring different jet multiplicities, that we have merged through the
MLM scheme~\cite{Mangano:2006rw,Alwall:2008qv}. The merging scale as been set,
for each point, to $Q^{\rm match} = M_{\tilde{t}}/4$~GeV for a {\sc MadGraph5}
{\tt xqcut} parameter set to 125~GeV. The A14 {\sc Pythia}
tune~\cite{ATL-PHYS-PUB-2014-021} has been used while showering and hadronizing
events with {\sc Pythia}~8, and the simulation of the ATLAS detector has been
achieved with the \textsc{Delphes}~3 program~\cite{deFavereau:2013fsa}, assuming
a $b$-tagging efficiency of 60\% for a $p_T$-dependent mistagging rate equal to
$0.1 + 0.000038*p_T$.

\begin{figure}[t]
  \centering
    \includegraphics[width=0.7\textwidth]{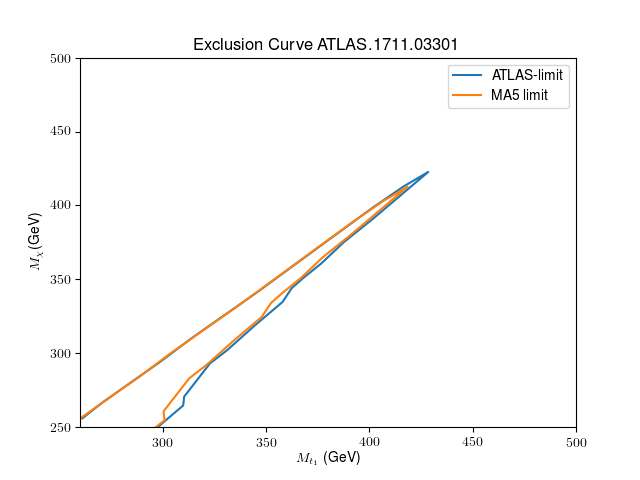}
  \caption{Exclusion contour in the $(M_{\tilde t}, M_{\tilde \chi})$ plane of
    the considered stop-neutralino class of simplified model. We compare the
    {\sc MadAnalysis}~5 findings (orange) with the official ATLAS numbers
    (blue).}
  \label{9-exclusion}
\end{figure}

In the absence of any official ATLAS cutflow for given benchmark scenarios, we
have decided to validate our reimplementation by reproducing the ATLAS exclusion
contour for a set of compressed benchmark points for which the stop decays as
$\tilde{t} _{1}\to c \chi_{1}^{0}$. Our results are presented in
Fig.~\ref{9-exclusion} in which we superimpose the exclusion contour obtained
with {\sc MadAnalysis}~5 (orange) with the official ATLAS one (blue). We observe
an excellent degree of agreement, which makes us considering our
reimplementation as validated.

\section{Summary}

We have implemented the ATLAS-EXOT-2016-27 analysis the \ma\ framework,
an analysis searching for dark matter models in the monojet channel and in
36.2~fb$^{-1}$ of ATLAS collision data at a center-of-mass energy of 13~TeV.
In the absence of any detailed validation material, we have validated our
reimplementation in reproducing the exclusion curve provided by ATLAS in the
context of a class of simplified models where the Standard Model is extended by
a neutralino and a stop that decays into the $\tilde{t}_1 \to c \chi_{1}^{0}$
channel. We have obtained an exceptionally good agreement, so that our
reimplementation has been considered as validated. It is available from \ma\
version 1.6 onwards, its Public Analysis Database and from {\sc InSpire}~\cite{%
1672234},\\
\hspace*{0.7cm}\url{http://doi.org/10.7484/INSPIREHEP.DATA.HUH5.239F}.

%% file: 3-atlas_1704_03848.tex
\chapter{ATLAS-EXOT-2016-32: an ATLAS monophoton analysis (36.1~fb$^{-1}$)}
\label{atlas_1704_03848}
{\it S.~Baek, T.~H~Jung}

\begin{abstract}
We present the \ma\ implementation and validation of the ATLAS-EXOT-2016-32
analysis, a search that targets a new physics signature featuring an energetic
photon and a large amount of missing transverse momentum. The results are
presented for an integrated luminosity of 36~fb$^{-1}$ of proton-proton
collisions at a center-of-mass energy of 13 TeV recorded by the ATLAS detector.
This analysis has been in particular designed to search for the pair
production of dark matter particles recoiling against a very energetic photon.
Our implementation has been validated by comparing our cutflow predictions with
those available from ATLAS.
\end{abstract}

\section{Introduction}
In this note, we summarize the \ma~\cite{Conte:2012fm,Conte:2014zja,%
Dumont:2014tja} implementation of the ATLAS search for the production of dark
matter in association with a hard photon~\cite{Aaboud:2017dor}. This search
focuses on 13~TeV LHC data and an integrated luminosity of 36.1~${\rm fb}^{-1}$,
and the details of this anlysis is documented on\\
  \hspace*{0.7cm}\url{https://atlas.web.cern.ch/Atlas/GROUPS/PHYSICS/PAPERS/EXOT-2016-32/}.

The typical dark matter models that are probed by such an analysis can be
embedded in the simplified model presented in Ref.~\cite{Backovic:2015soa}. In
this case, the Standard Model is supplemented by a Dirac fermionic dark matter
particle that can be produced in quark-antiquark annihilations via an
$s$-channel exchange of an axial-vector mediator. The corresponding Lagrangian
reads
\be
{\cal L}=g_{\chi} \bar X_D \gamma_\mu \gamma_5 X_D Y_1^\mu +
\sum_{i,j}\left[g^A_{d_{ij}}\bar d_i \gamma_\mu \gamma_5 d_j+g^A_{u_{ij}}\bar
u_i \gamma_\mu \gamma_5 u_j\right] \ ,
\ee
where $X_D$ denotes the fermionic dark matter candidate and $Y_1^\mu$ the
mediator. For simplicity, we ignore flavor-violating effects and consider flavor
universality, so that the new physics couplings satisfy
\be
  g^A_{d_{ij}}=g^A_{u_{ij}}=g_q \delta_{ij}\ ,
\ee
with $i,j=1, 2, 3$ being flavor indices. For the validation of our
reimplementation, we consider the benchmark scenario
defined in Ref.~\cite{Aaboud:2017dor} in which the universal coupling of the
mediator to quarks is set to $g_q=0.25$ and the mediator coupling to dark matter
is set to $g_\chi=1$. The new physics setup additionally includes a dark matter
mass of 10~GeV and a mediator mass of 800~GeV, which yields a mediator width of
44.01~GeV.

\section{Description of the implementation}

\subsection{Objects}
In the ATLAS-EXOT-2016-32 analysis, the signal region definition relies on
photons whose transverse energy $E_T^\gamma$ and pseudorapidity $\eta^\gamma$
satisfy
\be
  E_T^{\gamma}>10~{\rm GeV} \qquad\text{and}\qquad
   1.52<|\eta^\gamma|<2.37 \quad\text{or}\quad |\eta^\gamma|<1.37\ .
\ee
Their isolation is enforced by requiring that the sum $\Sigma_E$ of the energy
deposits in a cone of radius $\Delta R=0.4$ centered on the photon fullfils
\be
  \Sigma_E < 2.45~{\rm GeV} + 0.022 E_T^\gamma \ ,
\ee
and that the scalar sum $ \Sigma_{p_T}$ of the transverse momenta of the
non-conversion tracks lying in a cone of radius $\Delta R=0.2$ centered on the
photon satisfies
\be
  \Sigma_{p_T} < 0.05\times E_T^\gamma \ .
\ee

Electron candidates are required to have a transverse momentum $p_T^e$ and
pseudorapidity $\eta^e$ obeying to
\be
  p_T^e >7~{\rm GeV} \qquad\text{and}\qquad |\eta^e|<2.47\ ,
\ee
while the muon candidates are defined similarly,
\be
  p_T^\mu >6~{\rm GeV} \qquad\text{and}\qquad |\eta^\mu|<2.7\ .
\ee
Jets are recontructed by means of the anti-$k_T$
algorithm~\cite{Cacciari:2008gp}, with a radius parameter set to $R=0.4$,
and the analysis restricts itself to jet candidates with a transverse momentum
$p_T^j$ and pseudorapidity $\eta^j$ fullfilling
\be
  p_T^j>30~{\rm GeV} \qquad\text{and}\qquad |\eta|<4.5 \ .
\ee
The missing transverse momentum vector ${\bf E}_T^{\rm miss}$ is defined as the
opposite of the vector sum of the momenta of all reconstructed physics object
candidates, and the missing transverse energy is defined by the norm of this
vector,
\be
  E_T^{\rm miss}=|{\bf E}_T^{\rm miss}|\ .
\ee

\subsection{Event Selection}

Our reimplementation of the ATLAS monophoton search in \ma\ includes all five
signal regions described in the analysis (see the Table 2 in
Ref.~\cite{Aaboud:2017dor}). They all require to select events
featuring one hard photon with an energy
\be
  E_T^{\gamma}>150~{\rm GeV} , 
\ee
and well separated from the missing momentum in azimuth,
\be
  \Delta \phi(\gamma, {\bf E}_T^{\rm miss})>0.4 \ .
\ee
The missing energy significance is imposed to be large,
\be
  \frac{E_T^{\rm miss}}{\sqrt{\sum E_T}} > 8.5\,{\rm GeV}^{1/2} \ ,
\ee
and a (loose) jet veto is finally imposed. The selected events are hence allowed
to feature at most one jet that must be well separated from the
missing momentum in azimuth,
\be
  \Delta\phi(j, {\bf E}_T^{\rm miss})>0.4 \ .
\ee
The five signal regions are differentiated by means of different missing
energy selection criteria. Three inclusive regions SRI1, SRI2 and SRI3 are
respectively defined by imposing that
\be
  E_T^{\rm miss} > 150~{\rm GeV}, \qquad
  E_T^{\rm miss} > 225~{\rm GeV} \qquad\text{and}\qquad
  E_T^{\rm miss} > 300~{\rm GeV},
\ee
whilst two exclusive regions SRE1 and SRE2 focus on definite missing energy 
ranges,
\be
  E_T^{\rm miss} \in [150, 225]~{\rm GeV} \qquad\text{and}\qquad
  E_T^{\rm miss} \in [225, 300]~{\rm GeV}.
\ee
The provided validation material is however only available for the SRI1
region~\cite{Aaboud:2017dor}.

\section{Validation}
\subsection{Event Generation}
In order to validate our reimplementation of the ATLAS analysis, we focus on the
simplified model introduced above. In order to generate hard scattering signal
events, we use the UFO~\cite{Degrande:2011ua} model associated with the
considered simplified dark matter model~\cite{Backovic:2015soa} that has been
generated with the {\sc FeynRules}~\cite{Alloul:2013bka} and
NLOCT~\cite{Degrande:2014vpa} programs. We have imported this
model into {\sc MadGraph5\_aMC@NLO} version~2.6.0~\cite{Alwall:2014hca} and
generated parton-level events by convoluting matrix elements at the
next-to-leading order (NLO) accuracy in QCD with the NLO set of NNPDF~3.0 parton
distribution functions~\cite{Ball:2014uwa}. Those events have then been showered
and hadronized within the {\sc Pythia 8.2} environment~\cite{Sjostrand:2014zea},
and the simulation of the detector response has been made with
{\sc Delphes~3}~\cite{deFavereau:2013fsa} that internally relies on {\sc
FastJet}~\cite{Cacciari:2011ma} for object reconstruction. We have used our \ma\
reimplementation to calculate the signal selection efficiencies.

\subsection{Comparison with the official results}
In Table.~\ref{table:cutflow}, we compare the results obtained with our
implementation to the official numbers provided by the ATLAS collaboration. The
discrepancy is characterized according to the measure
\be\label{eq:error}
  |{\rm error}| = 
    \left|\frac{{\rm MA5}-{\rm Official}}{{\rm Official}}\right| \ .
\ee
We observe that the disagreement, on a cut-by-cut basis, is of at most 20\%, and
even smaller than that for most cuts. We
therefore consider our analyssis as validated.

\begin{table}
\centering
 \renewcommand{\arraystretch}{1.4}
\begin{tabular}{l|l|l|l}
  cuts                                                              & MA5              & Official        & error      \\ \hline
Initial                                                           & 1198             & 1198            &            \\ 
$\met>150$ GeV                                                    & 882.1($-26.37\%$) & 736($-38.56\%$) & $19.85\%$  \\ 
$p_T^{\gamma 1}>150$ GeV and $|\eta|<2.37$                     & 683.1($-22.56\%$) & 700($-4.89\%$)  & $-2.41\%$ \\ 
Tight leading photon                                              & 570.0($-16.56\%$) & 658($-6.00\%$)  & $-13.38\%$ \\ 
$\Delta \phi(\gamma, \met)>0.4$                                   & 568.6($-0.24\%$)  & 620($-5.78\%$)  & $-8.30\%$   \\ 
$\met/\sqrt{\sum E_T}>8.5~{\rm GeV}^{1/2}$                        & 555.4($-2.32\%$)  & 596($-3.87\%$)  & $-6.81\%$   \\ 
$N_{\rm jet}<2$ and $\Delta \phi({\rm jet}, \met)$\textgreater0.4 & 447.6($-17.13\%$) & 461($-22.65\%$) & $-2.91\%$   \\ 
Lepton veto                                                       & 447.6($-0.00\%$)  & 460($-0.21\%$)  & $-2.7\%$   \\
\end{tabular}
\caption{Comparison of the cutflow predicted by \ma\ with the one provided by
  the ATLAS collaboration.}
\label{table:cutflow}
\end{table}

\section{Summary}
We have implemented in \ma\ the five signal regions of the ATLAS monophoton
analysis of 36.1~fb$^{-1}$ of LHC collision data at a center-of-mass energy of
13~TeV. We have validated our implementation in the context of a Dirac
fermionic dark matter simplified model featuring an axial-vector mediator by
comparing our predictions for the cutflow with the official
one provided by ATLAS in Ref.~\cite{Aaboud:2017dor}. We have found an
agreement that is better than at the 20\% level, so that we consider our
reimplementation as validated. It is available from \ma\
version 1.6 onwards, its Public Analysis Database and from {\sc InSpire}~\cite{%
1642639},\\
\hspace*{0.7cm}\url{http://doi.org/10.7484/INSPIREHEP.DATA.88NC.0FER.1}.

%% file: 2-cms_exo_16_012.tex
\chapter{CMS-EXO-16-012: a CMS mono-Higgs analysis (3.2~fb$^{-1}$)}
\label{cms-exo-16-012}
{\it S.~Ahn, J.~Park and W.~Zhang}

\begin{abstract}
We present the implementation and validation of the CMS-EXO-16-012 analysis
within \ma. This search targets events featuring a large missing transverse
momentum and the signature of a Higgs boson decaying into a pair of bottom
quarks or photons, and focuses on 2.3 fb$^{-1}$ of proton-proton collisions at a
center-of-mass energy of 13 TeV. In our reimplementation, we only focus on the
$\gamma\gamma$ final state and validate our reimplementation in the context of a
two-Higgs-doublet model including an extra neutral gauge boson.
\end{abstract}

\section{Introduction}
In this document, we detail the \ma~\cite{Conte:2012fm,Conte:2014zja,%
Dumont:2014tja} implementation of the CMS search for the associated production
of dark matter with a Higgs boson decaying into a $b \bar b$ or $\gamma\gamma$
pair. This
search focuses on the analysis of 2.3~fb$^{-1}$ of proton-proton collision data
at a center-of-mass energy of $\sqrt{s} = 13$~TeV~\cite{Sirunyan:2017hnk}.
The $b\bar{b}$ channel subanalysis is divided in two regimes, {\it i.e.} a
resolved
regime where the Higgs boson decays into two distinct reconstruced $b$-jets, and
a Lorentz-boosted regime where the Higgs boson is reconstructed as a single fat
jet. In this last case, the signal extraction is performed through a
simultaneous fit of signal regions and background-enriched control regions. We
have not been able to reproduce this fit consequently to the lack of associated
public information, and we have therefore not reimplemented this analysis
strategy. On the other hand, the $\gamma\gamma$ channel search is
performed by seeking an excess of events over the Standard Model expectation
in the diphoton mass spectrum, which solely relies on a cut-and-count approach.

\begin{figure}[t]
    \centering
    \includegraphics[width=0.2\textwidth]{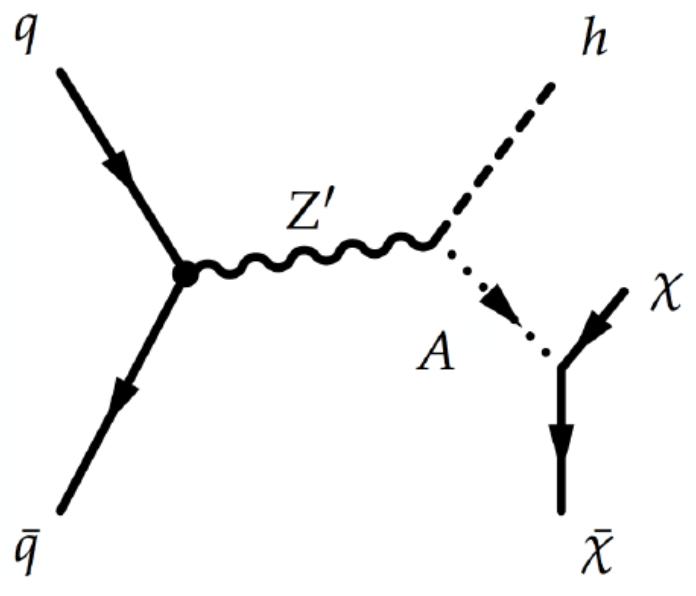}
    \caption{Leading order Feynman diagram yielding the production of the signal
    of interest in the considered $Z'$--2HDM simplified model. The associated
    signature consists of a Higgs boson produced in association with
    missing transverse momentum.}
    \label{fig:model1}
\end{figure}

The analysis presented in Ref.~\cite{Sirunyan:2017hnk} has been interpreted
using a benchmark simplified model in which a two-Higgs-doublet model is
supplemented by an extra $Z'$ boson and a dark matter particle $\chi$
($Z'$-2HDM)~\cite{Berlin:2014cfa,%
Abercrombie:2015wmb}. The signal that is probed by the analysis corresponds to
the resonant production of a heavy $Z'$ vector boson which further decays into a
Standard-Model-like Higgs boson $h$ and an intermediate heavy pseudoscalar boson
$A$ that connects the visible sector to a dark sector. The mediator $A$ hence
decays into a pair of dark matter particles. The entire process,
\be
 p p \to Z' \to h A \to h\ \bar\chi \chi \ ,
\ee
is described
in Fig.~\ref{fig:model1}. However, this signature is quite generic and its
reimplementation within the \ma~framework could enable more reinterpretations.
For example it could be used to probe other scalar extensions
of the Standard Model, noteworthy in a more general two-Higgs-doublet plus
singlet extensions of the Standard Model or in a supersymmetric context. In
particular, such as signature could provide an interesting handle on the NMSSM,
where the $Z' A h$ coupling is replaced by a $ A_1 A_2 h$ or
$h_3 h_2 h_1$ interaction with $A_{1,2}$ and $h_{1,2,3}$ respectively being
$CP$-odd and $CP$-even scalars~\cite{Baum:2017gbj}.

\section{Description of the analysis}
To enforce the compatibility with the presence of a Higgs boson decaying into
two photons, events are selected if they feature a photon pair satisfying given
invariant mass and transverse momentum ($p_T$) requirements. Moreover, fake
photons are rejected through constraints on the calorimetric activity of the
reconstructed photons and their isolation. The signal region is further defined
by imposing constraints on the ratio of the photon $p_T$ to the diphoton
invariant-mass, as well as on the missing transverse momentum
and on the angular separation between the reconstructed Higgs boson and the
missing momentum.

\subsection{Objects definition and preselection}
In this analysis, photons are identified following different ways. A cut-based
identification is first performed, relying on a loose working point. The exact
selections are presented in Ref.~\cite{Khachatryan:2015iwa}, as well as in the
CMS-PAS-EXO-16-012 analysis note~\cite{Sirunyan:2017hnk}. In practice, isolation
is imposed by restricting the calorimetric activity in a cone of radius
$\Delta R = 0.3$ centered on the photon through three variables, $I_\pm$, $I_0$
and $I_\gamma$. These respectively correspond to the amount of calorimetric
deposits originating from charged hadrons, neutral hadrons and photons lying in
the considered cone.

The signal region is defined by requiring the presence of two photons whose
transverse momenta fulfill
\be
  p_T(\gamma_1) > 30~{\rm GeV}  \qquad\text{and}\qquad
  p_T(\gamma_2) > 18~{\rm GeV}.
\ee
Fake photons are rejected by requiring that the ratio of the amount of energy
deposited in the hadronic calorimeter is of at most 10\% of the amount of energy
deposited on the electromagnetic calorimeter,
\be
  H/E < 0.1 \ ,
\ee
and photon isolation is ensured by the selections on the  $I_\pm$, $I_0$ and
$I_\gamma$ variables given in Table~\ref{tab:2-photonI}. Whilst the isolation
requirement related to the neutral particles should include the so-called $\rho$
correction that accounts for the dependence of the pileup transverse energy
density on the photon pseudorapidity, $\rho$ being the median of the transverse
energy density per unit area, we ignore this correction in our implementation
due to the lack of relevant information.

Events are finally further preselected by requiring that the invariant mass of
the diphoton system satisfies
\be
  m_{\gamma\gamma} > 95~{\rm GeV} \ ,
\ee
in order to be compatible with the decay of a Higgs boson.

\begin{table}
\begin{center}
   \renewcommand{\arraystretch}{1.4}
   \begin{tabular}{c | c c}
   Variable & Barrel & Endcap \\
   \hline
   $I_\pm$ [GeV]& $<$ 3.32& $<$ 1.97\\
   $I_0$ [GeV]   & $< 1.92 + 0.14 p_T^\gamma + 0.000019(p_T^\gamma)^2$ &
      $< 11.86 + 0.0139 p_T^\gamma + 0.000025(p_T^\gamma)^2$\\
   $I_\gamma$ [GeV]& $<0.81 + 0.0053 p_T^\gamma$& $< 0.83 + 0.0034 p_T^\gamma$\\
   \end{tabular}
   \caption{Requirements imposed on the photon isolation. We distinguish photons
     reconstructed in the barrel (second column) and in the endcap (third
     column), and $p_T^\gamma$ denotes the photon transverse momentum.}
   \label{tab:2-photonI}
\end{center}
\end{table}

\subsection{Signal selections}

After the preselection described above, the CMS-PAS-EXO-16-012 analysis includes
a series of cuts defining the signal region. These kinematic selections consist
of additional constraints on the $p_T$ of the two photons,
\be
  \frac{p_T(\gamma_1)}{m_{\gamma\gamma}} > 0.5
  \qquad\text{and}\qquad
  \frac{p_T(\gamma_2)}{m_{\gamma\gamma}} > 0.25\ ,
\ee
for the leading and next-to-leading photon respectively, and of a selection on
the diphoton transverse momentum and on the missing transverse energy
$E_T^{\rm miss}$,
\be
  p_{T_{\gamma\gamma}} > 90~{\rm  GeV}
  \qquad\text{and}\qquad
  E_T^{\rm miss} > 105~{\rm GeV}.
\ee
Two extra cuts further constrain the angular seperation between the
missing transverse momentum $\mathbf{p}_T^{\rm miss}$ and the diphoton system,
\be
  |\Delta\phi(\gamma\gamma,\, \mathbf{p}_T^{\rm miss})| > 2.1
  \qquad\text{and}\qquad
  \min_j(|\Delta\phi(j,\,\mathbf{p}_T^{\rm miss})|) > 0.5 \ ,
\ee
where the minimization has to account for all jets with a transverse momentum
larger
than 50~GeV. In this analysis, jets are recontructed by means of the anti-$k_T$
algorithm~\cite{Cacciari:2008gp}, with a radius parameter set to $R=0.4$.
Finally the diphoton invariant mass is further imposed to satisfy
\be
   120~{\rm GeV} < m_{\gamma\gamma} < 130~{\rm GeV} \ .
\ee

\section{Validation}

In order to validate our reimplementation, we focus on the $Z'$--2HDM model
described
above and on the production of a heavy $Z'$ boson that decays into a Higgs boson
and a pair of dark matter particles via an intermediate pseudoscalar state $A$
(see Fig.~\ref{fig:model1} for a representative Feynman diagram).
Hard-scattering signal events are generated with
\textsc{MadGraph5}\_aMC@NLO~\cite{Alwall:2014hca}, the matrix elements being
generated from the model information provided through an appropriate
UFO~\cite{Degrande:2011ua} model shared by CMS and convoluted with the
next-to-leading-order set of NNPDF~3.0 parton densities~\cite{Ball:2014uwa}.
Our tests focus on several benchmark scenarios featuring each a different
$Z'$-boson mass $M_{Z'}$. The simulation of the hadronic environment (parton
showering and hadronization) is performed by means of
\textsc{Pythia 8}~\cite{Sjostrand:2014zea}, that is also used to handle the
decay of the final-state Higgs boson. The simulation of the response of the
CMS detector is achieved via \textsc{Delphes 3}~\cite{deFavereau:2013fsa}, that
internally relies on {\sc FastJet}~\cite{Cacciari:2011ma} for object
reconstruction, with an tuned detector configuration including updated
$b$-tagging and reconstruction performances.

We make use of our reimplementation of the CMS-PAS-EXO-16-012 analysis to
compute \ma\ predictions for the acceptance times efficiency values for the
different scenarios. Our reimplementation is then validated by comparing our
results with the official numbers from CMS.

\subsection{Event Generation}
Hard scattering events are generated by making use of the
\textsc{MadGraph5}\_aMC@NLO package, together with the UFO model available on
the CMS public repository,\\
  \hspace*{.05cm}\url{http://rkhurana.web.cern.ch/rkhurana/monoH/models/}\\
The necessary configuration files for each of the considered benchmarks can be
found from the {\sc MadGraph5} generator repository of CMS,\\
  \hspace*{.05cm}\url{https://github.com/cms-sw/genproductions/tree/mg240/bin/MadGraph5_aMCatNLO}\\
in the folder\\
  \hspace*{.05cm}\url{cards/production/13TeV/monoHiggs/Zp2HDM/Zprime_A0h_A0chichi}

We fix the masses of the pseudoscalar state and of the dark matter particle to
300~GeV and 100~GeV, respectively, and set the decay width of the pseudoscalar
to 8.95~GeV. We investigate several configurations for the properties of the
$Z'$ boson. Its mass is hence varied and fixed to 600, 800, 1000, 1200, 1400,
1700, 2000 and 2500 GeV for the different setups. All the $Z'$ couplings to
Standard Model particles $g_{\rm SM}$ are chosen to be equal to 0.8, while the
coupling to dark matter is fixed to 1~\cite{Abercrombie:2015wmb}. The
corresponding $Z'$-boson width for each mass value is given in
Table~\ref{tab:2-Zpwidth}.
\begin{table}
\begin{center}
   \renewcommand{\arraystretch}{1.4}
   \begin{tabular}{c || c | c | c | c | c | c | c | c |}
   $M_{Z'}$ (GeV) & 600 & 800 & 1000 & 1200 & 1400 & 1700 & 2000 & 2500 \\
    \hline
    $\Gamma_{Z'}$ (GeV) & 11.223 & 15.765 & 20.225 & 24.624 & 28.982
& 35.473 & 41.927 & 52.639 \\
    \end{tabular}
  \caption{\label{tab:2-Zpwidth}Values of the $Z'$ total width for each
    benchmark point used in  the validation process.}
\end{center}
\end{table}

We enforce the Higgs boson to decay into a diphoton system by setting
appropriately the {\sc Pythia}~8 configuration. This requires to modify two
{\sc Pythia 8} input files,
\verb+Pythia8CUEP8M1Settings_cfi.py+ and \verb+Pythia8CommonSettings_cfi.py+,
which we have been again found on public repositories of the CMS generator
group,\\
  \hspace*{.05cm}\url{https://github.com/cms-sw/cmssw/tree/CMSSW_7_1_9_patch}\\
  \hspace*{.05cm}\url{https://github.com/cms-sw/cmssw/tree/CMSSW_7_2_X}\\
respectively, in the \verb+Configuration/Generator/python+ subfolder in both
cases.

Concerning the simulation of the CMS detector, we have slightly modified the
configuration that has been designed for the reimplementation of the
CMS-EXO-16-037 analysis and that is available on\\
  \hspace*{.05cm}\url{http://madanalysis.irmp.ucl.ac.be/wiki/PublicAnalysisDatabase}\\
Compared with the default settings, the $b$-tagging and lepton and photon
reconstruction performances have been updated according to
Refs.~\cite{CMS:2016kkf,Khachatryan:2015iwa}. In particular, we make use of the
cMVAv2 loose $b$-tagging working point, which corresponding to a correct
$b$-tagging efficiency of about 83\% for a misidentification probability of
about 10\%. We have also defined the dark matter particle as an invisible state
that does not deposit energy in the calorimeters.

 \begin{table}
 \begin{center}
 \renewcommand{\arraystretch}{1.4}
 \begin{tabular}{c c c c}
   \multicolumn{4}{c}{Acceptance $\times$ efficiency $(A \cdot \epsilon)$}\\
   \hline
    $m_{Z'}$ (GeV)& CMS EXO-16-012& MA5& Difference\\
    \hline
       600& 0.317 $\pm$ 0.004& 0.355 $\pm$ 0.001& -11 \%\\
       800& 0.399 $\pm$ 0.004& 0.451 $\pm$ 0.001& -13 \%\\
       1000& 0.444 $\pm$ 0.004& 0.494 $\pm$ 0.001& -8.2 \%\\
       1200& 0.474 $\pm$ 0.004& 0.513 $\pm$ 0.001& -0.6 \%\\
       1400& 0.492 $\pm$ 0.004& 0.515  $\pm$ 0.001& -4.7 \%\\
       1700& 0.493 $\pm$ 0.004& 0.494 $\pm$ 0.001& -0.2 \%\\
       2000& 0.351 $\pm$ 0.004& 0.355 $\pm$ 0.001& -1.1 \%\\
       2500& 0.213 $\pm$ 0.004& 0.208 $\pm$ 0.001& 2.3 \%\\
  \end{tabular}
  \caption{Comparison of the signal acceptance times efficiencies predictions
    made by \ma\ with the CMS official numbers. The difference is calculated
    according to Eq.~\eqref{eq:2-delta}.}
  \label{tab:2-results}
\end{center}
\end{table}

\subsection{Comparision with official results}
As CMS has not provided detailed validation information, we have validated our
implementation on the basis of the available material. We present the product of
signal acceptance and selection efficiency for each considered $Z'$ mass point,
and we define the difference with the official numbers as 
\be
 \delta =  1-\frac{(A\cdot\epsilon)^{\rm MA5}}{(A\cdot\epsilon)^{\rm CMS}} \ ,
\label{eq:2-delta}\ee
The results are given in Table~\ref{tab:2-results}.

\begin{figure}
  \centering
  \includegraphics[width=0.48\columnwidth]{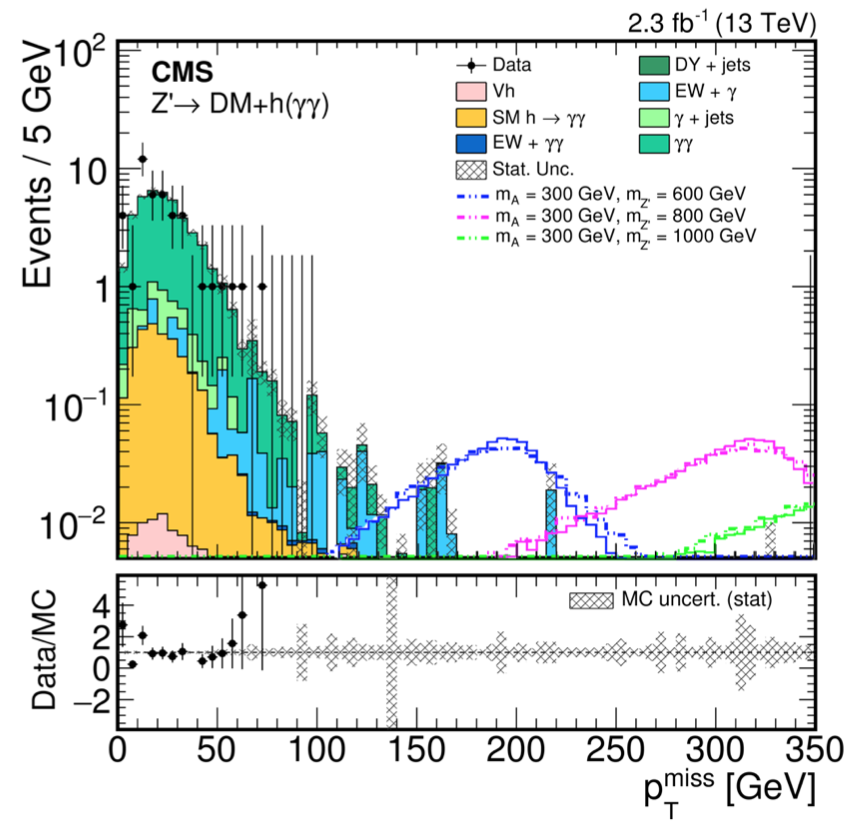}
  \includegraphics[width=0.49\columnwidth]{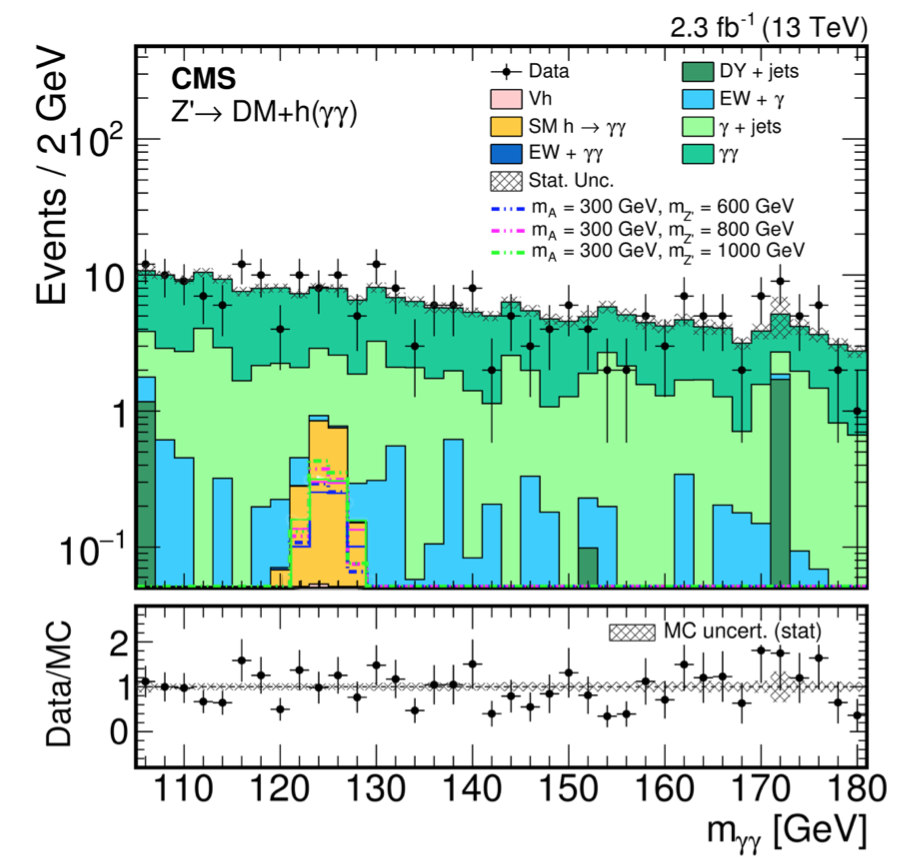}
  \caption{Missing transverse energy (left) and diphoton invariant mass (right)
    distributions after all selection criteria have been imposed, except the
    one on the missing enegy (both cases) and the one of the diphoton invariant
    mass (right panel only). The dotted lines are the official CMS results
    taken from Ref.~\cite{Sirunyan:2017hnk} and the solid lines are the \ma\
    predictions.}
  \label{fig:2-distri}
\end{figure}

Moreover, we present, for representative signal scenarios, the missing
transverse energy and diphoton invariant mass distributions in
Fig.~\ref{fig:2-distri} after normalizing our signal distributions similarly to
CMS. For all performed tests, a good agreement is obtained.

\section{Summary}

In this note, we reported the \ma\ reimplementation of the CMS-EXO-16-012
and analysis and its validation. We compared signal selection efficicies times
acceptance for varied benchmark scenarios, as well as two differential
distributions. An overal agreement has been found, the differences being of at
most 13\%. This analysis is thus considered as validated and has been made
available from \ma\ version 1.6 onwards, its Public Analysis Database and from
{\sc InSpire}~\cite{1642631},\\
\hspace*{0.7cm}\url{http://doi.org/10.7484/INSPIREHEP.DATA.JT56.DDC3.1}.

%% file: 5-cms_exo_16_022.tex
\chapter{CMS-EXO-16-022: a CMS long-lived lepton analysis (2.6~fb$^{-1}$)}

{\it Jung Chang}

\begin{abstract}
We present the \ma\ implementation and validation of the CMS-EXO-2016-22
analysis, which documents a search for new long-lived particles that decay into
electrons and muons. The results are based on a dataset of proton-proton
collisions recorded by CMS with a center-of-mass energy of 13~TeV and an
integrated luminosity of 2.6~fb$^{-1}$. The validation of our reimplementation
is based on a comparison of the expected number of signal event counts in the
signal regions with information provided by the CMS collaboration, with signal
events corresponding to a benchmark model featuring pair-produced long-lived
top squarks.
\end{abstract}

\section{Introduction}
In this contribution, we summarize the \ma~\cite{Conte:2012fm, Conte:2014zja,%
Dumont:2014tja} implementation of the CMS-EXO-16-022 analysis, a search for
long-lived particles in 2.6~fb$^{-1}$ of LHC proton-proton collision
data at a center-of-mass energy of 13~TeV~\cite{CMS:2016isf}, that we present
together with its validation. The simulation of the signal events used for the
validation relies on a \ma\ tune of {\sc Delphes 3}~\cite{deFavereau:2013fsa}
that has been specifically designed to deal with long-lived particles. It in
particular allows for handling neutral long-lived particles that decay into
leptons within the volume of the tracker. Reconstruction efficiencies can be
applied to displaced tracks and various related parameters can be accessed at
the analysis level by means of a dedicated \ma\ version.

In practice, the simulation of the displaced leptons is performed through
efficiencies and resolution functions that the user can specify in the
{\sc Delphes} card. More information is available on the web page\\
\hspace*{0.5cm} \url{
  https://madanalysis.irmp.ucl.ac.be/wiki/MA5LongLivedParticle}\\
that also includes a download link to the special version of \ma\ that has to
be employed. We have used the reconstruction efficiency depending on the impact
parameter $d_0$ provided in Ref.~\cite{Khachatryan:2014mea}.

For our validation, we have focused on an $R$-parity-violating (RPV)
supersymmetric
scenario featuring a long-lived stop. Relying on the material provided by the
CMS collaboration, we have considered four different stop decay lengths fixed to
\be
  c\tau_{\tilde t} = 0.1,\ 1,\ 10 \ \ \text{and}\ \ 100~{\rm  cm},
\ee
respectively, for a stop mass of $m_{\tilde{t}}=700$~GeV in all cases. The
stop is then assumed to decay via an RPV channel,
\be
  \tilde{t}\rightarrow b \ell \qquad\text{with}\qquad
  \ell=e \ \ \text{or}\ \ \mu\ .
\ee
For simplicity, lepton universality has been assumed, so that the stop
branching fraction into an electron, muon and tau final state equals 1/3 in all
cases. The benchmark information corresponds to the Snowmass Points and Slopes
scenario SPS1a~\cite{Allanach:2002nj} that has been provided by the CMS
collaboration.

We have made use of our reimplementation of the CMS-EXO-16-022 analysis to
compute \ma\ predictions for the expected number of signal events in the
different signal regions defined in the CMS analysis. This has allowed us to
validate our reimplementation by comparing our predictions with the official
numbers from CMS.

\section{Description of the analysis}
As mentioned above, the CMS-EXO-16-022 analysis investigate new physics in a
channel where two displaced leptons, with a transverse impact parameter lying
between 200~$\mu$m and 10~cm, are observed. This analysis is particularly
sensitive to RPV supersymmetric signals as they could originate from the
production of a pair of long-lived top squarks that decay into a lepton and a
$b$-jet. While any combination of leptons is theoretically allowed, the analysis
focuses on the production of one muon and one electron only.

\subsection{Object definition and preselection}
The analysis preselects events that feature exactly one electron and one muon
that are well reconstructed and isolated. Selected events must have passed a
dedicated trigger targeting displaced electron-muon pairs where both leptons
have a transverse momentum $p_T^\ell$ satisfying
\be
  p_T^\ell > 38~{\rm GeV}.
\ee
Both leptons are then required to be central, with a pseudorapidity $\eta^\ell$
fulfilling
\be
  |\eta^\ell|<2.4 \ ,
\ee
and with a transverse momentum constrained to satisfy
\be
  p_T^e > 42~{\rm GeV} \qquad\text{and}\qquad
  p_T^\mu > 40~{\rm GeV}
\ee
for electrons and muons respectively. Moreover, both leptons are required to be
well separated from each other, in the transverse plane,
\be
  \Delta R(e,\mu) > 0.5\ ,
\ee
and are required to satisfy the isolation requirements
\be
  \frac{1}{p_{T}}\sum_{i} (p_{\text{T}})_i <
  \begin{cases}
     0.065 & \text{for }\ell=\text{e with } 1.57<|\eta^e|<2.4\\
     0.035 & \text{for }\ell=\text{e with } |\eta^e|<1.44\\
     0.015 & \text{for }\ell=\mu
  \end{cases} \ ,
\ee
where the sum is considered over all reconstructed particles within a $\Delta R$
cone of 0.3 (electrons) or 0.4 (muons), and where the lepton candidate
itself is excluded from the sum. Additionally, the lepton candidates are
required to originate from the pixel detector, which is achieved by imposing a
threshold on the transverse impact parameter $d_0^\ell$,
\be
  d_0^\ell < 10~{\rm cm}.
\ee

\subsection{Signal region selections}
The analysis contains three signal search regions whose definition varies
according to the values of the transverse impact parameters $d_0^\ell$ of the
two
leptons. The tight search region (SR III) requires both leptons to be displaced
by more than 10~cm,
\be {\rm SR~III:}\quad\quad
  d_0^{\ell_1} > 1000~\mu{\rm m} \qquad\text{and}\qquad
  d_0^{\ell_2} > 1000~\mu{\rm m}  \ ,
\ee
while an intermediate signal region SR~II allows for smaller displacements,
\be {\rm SR~II:}\quad\quad
  d_0^{\ell_1} > 500~\mu{\rm m} \qquad\text{and}\qquad
  d_0^{\ell_2} > 500~\mu{\rm m}  \ .
\ee
Finally, a looser signal region SR~I allows for even smaller displaced leptons,
featuring
\be {\rm SR~I:}\quad\quad
  d_0^{\ell_1} > 200~\mu{\rm m} \qquad\text{and}\qquad
  d_0^{\ell_2} > 200~\mu{\rm m}  \ .
\ee
Overlaps are removed from the signal regions by explicitly excluding the
tighter signal regions from the looser. For example, events populating the
SR~III region are excluded from the SR~II and SR~I regions, and events
populating the SR~II region are not allowed to populate the SR~I region.

\section{Validation}

\subsection{Event Generation}
In order to validate the CMS-EXO-16-022 \ma\ reimplementation, we focus on the
SPS1a supersymmetric
scenario whose parameterization has been provided by the CMS collaboration under
the form of an appropriate SLHA file~\cite{Skands:2003cj}. The stop decay table,
mass and width have been modified according to the requirement of the considered
benchmark scenarios.

Event generation relies on \textsc{Pythia8} (v~8.226)~\cite{Sjostrand:2014zea},
after making use of the command card provided by the CMS collaboration. This
corresponds to the {\sc Pythia} script,
\begin{verbatim}
  SUSY:gg2squarkantisquark  = on
  SUSY:qqbar2squarkantisquark= on
  SLHA:useDecayTable = true
  RHadrons:allow  = on
  1000006:tau0 = 1000 !mm
\end{verbatim}
in which we have turned on the {\tt RHadrons} command to enable stop
hadronization and the {\tt tau0} attribute of the particle class to set the stop
width.

We reweight our events so that the total production rate for stop
pair-production in proton-proton collisions at a center-of-mass energy of 13~TeV
matches the NLO+NLL predictions~\cite{Borschensky:2014cia},
\be
  \sigma(p\ p\to \tilde t\ \tilde t^\dag)\Big|_{m_{\tilde t} = 700~{\rm GeV}} = 
   0.067~{\rm pb}.
\ee
The event weight moreover includes a normalization factor accounting for
an integrated luminosity of 2.6~fb$^{-1}$.

\begin{figure}
  \centering
    \includegraphics[width=0.35\textwidth]{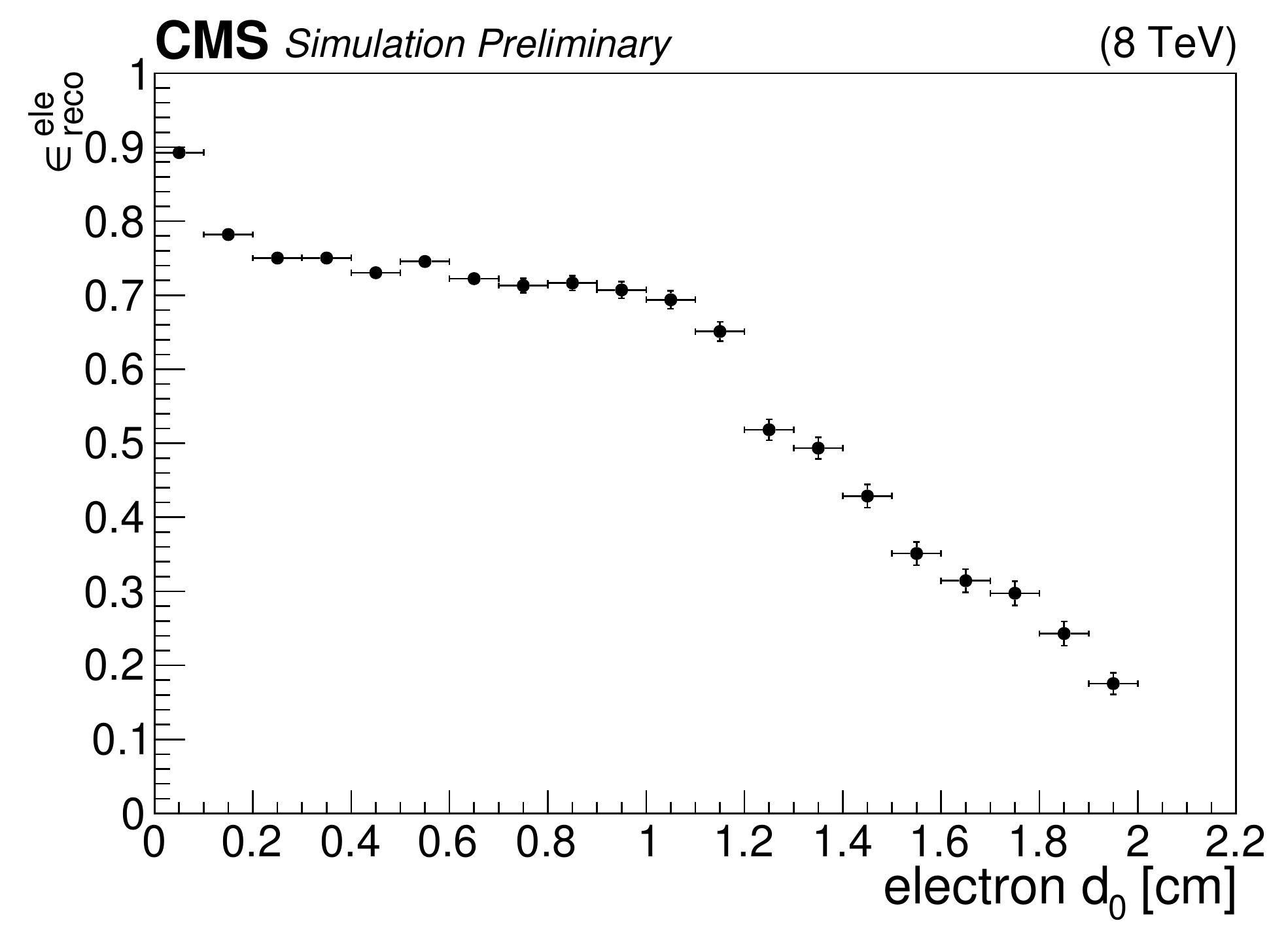}
    \includegraphics[width=0.35\textwidth]{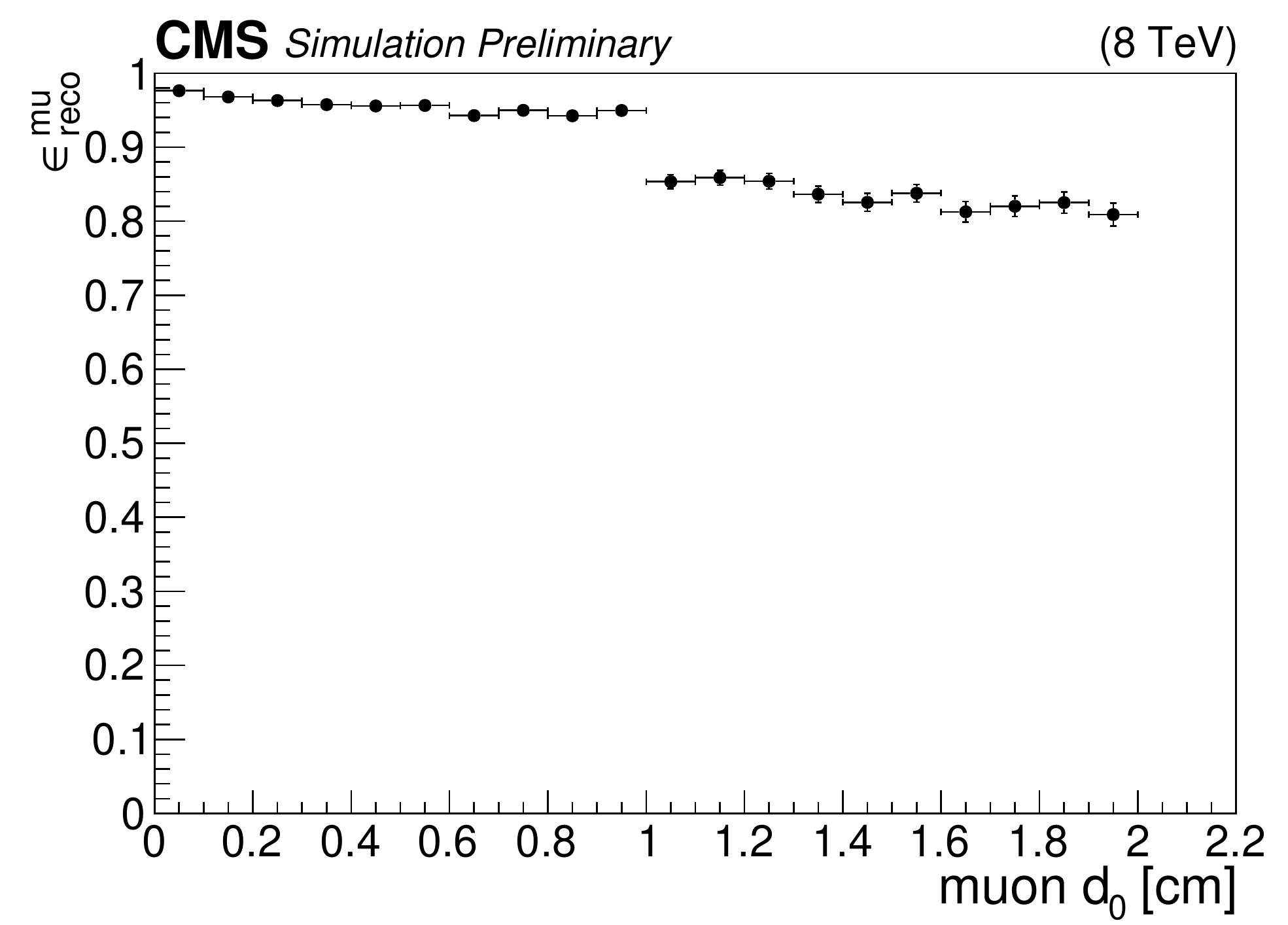}
    \includegraphics[width=0.35\textwidth]{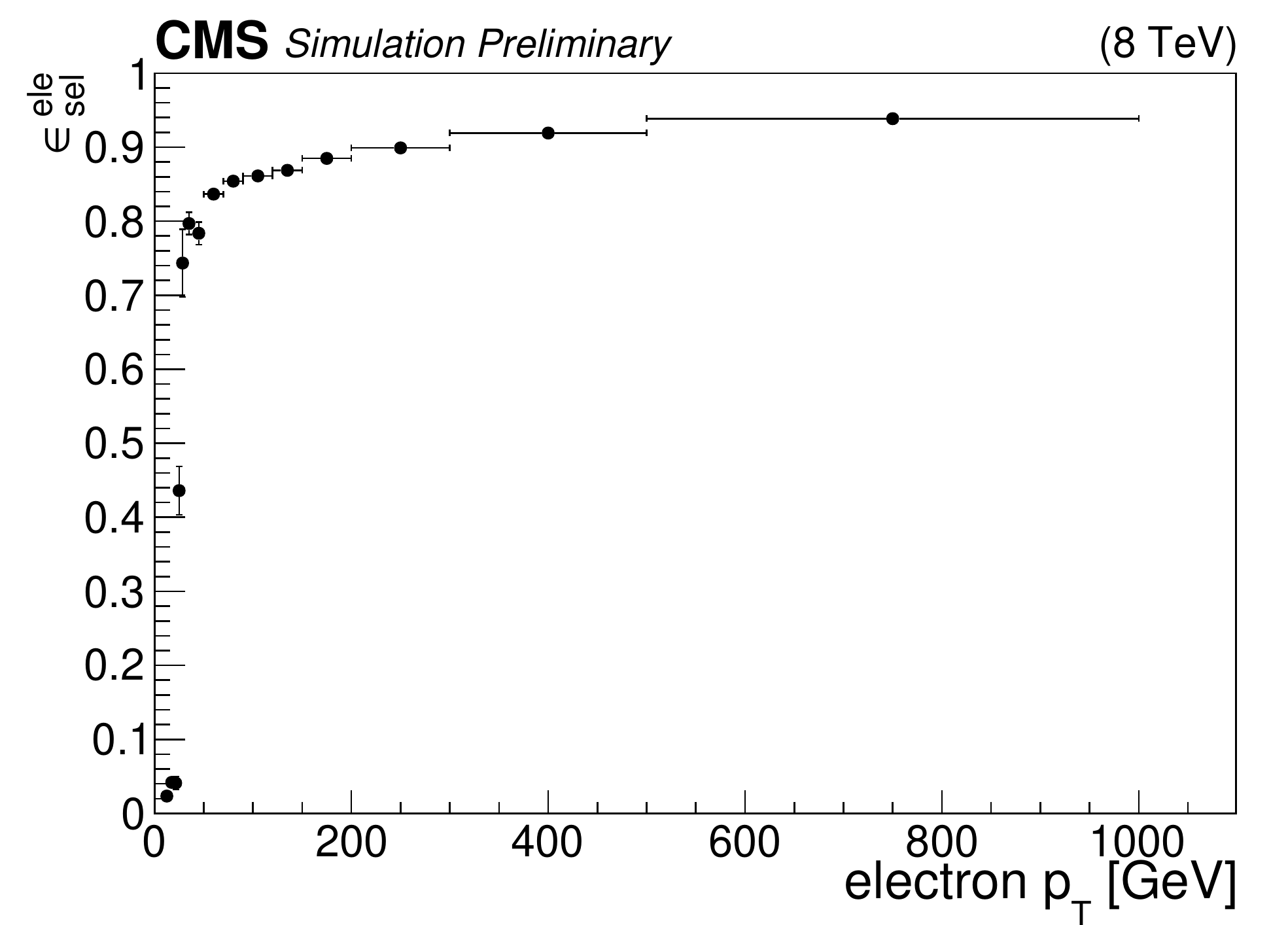}
    \includegraphics[width=0.35\textwidth]{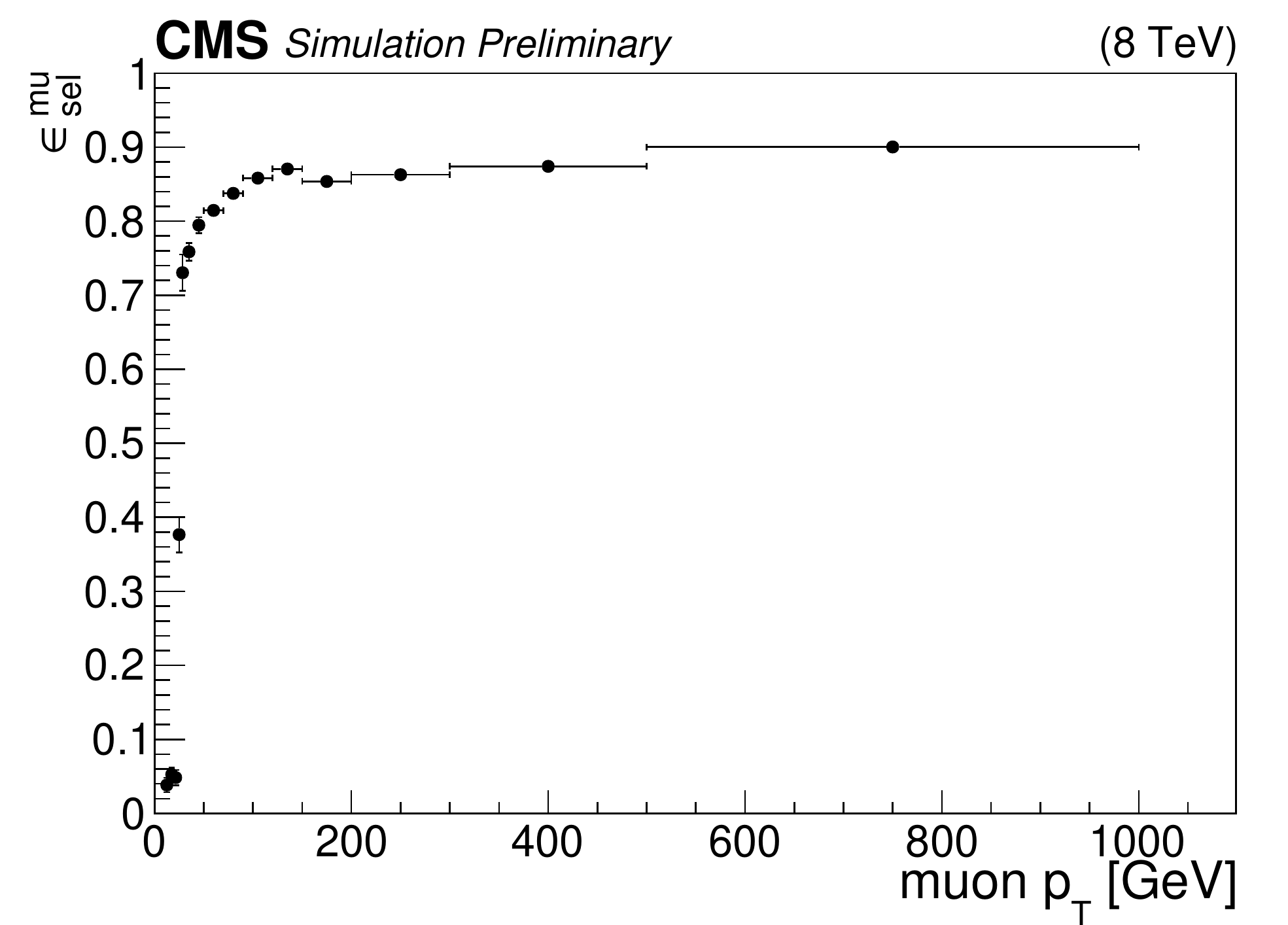}
  \caption{Reconstruction (upper panels) and selection (lower panels)
    efficiencies associated with displaced electrons and muons, as provided on
    \url{
    https://twiki.cern.ch/CMSPublic/DisplacedSusyParametrisationStudyForUser}.
  }\label{fig:5-efficiencies}
\end{figure}

The simulation of the response of the detector is achieved via the
\textsc{Delphes 3}~\cite{deFavereau:2013fsa} program and its internal use of
{\sc FastJet}~\cite{Cacciari:2011ma} for object reconstruction. Our detector
simulation includes reconstruction and selection efficiencies for displaced
electrons and muons, as provided on the public CMS webpage\\
\hspace*{0.5cm} \url{
  https://twiki.cern.ch/CMSPublic/DisplacedSusyParametrisationStudyForUser}\\
and presented on Figure~\ref{fig:5-efficiencies}.

\subsection{Comparision with official results}

\begin{table}
\renewcommand{\arraystretch}{1.4}
\begin{center}
  \begin{tabular}{l || l | c  c | c }
    Region & $c\tau_{\tilde t}$ [cm] & MA5 & CMS& Difference [\%] \\ \hline\hline
    \multirow{4}{*}{\bf SR-I}& 0.1 & 3.89   & 3.8  & 2.30\\
                             & 1   & 4.44   & 5.2  & 14.51\\
                             & 10  & 0.697  & 0.8  & 12.84\\
                             & 100 & 0.0610 & 0.009& $> 100\%$\\ \hline
    \multirow{4}{*}{\bf SR-II}& 0.1 & 0.924  & 0.94 & 1.71\\
                             & 1   & 3.87   & 4.1  & 5.61\\
                             & 10  & 0.854  & 1.0  & 14.58\\
                             & 100 & 0.0662 & 0.03 & $\sim 100\%$\\ \hline
    \multirow{4}{*}{\bf SR-III}& 0.1 & 0.139  & 0.16 & 12.84\\
                             & 1   & 6.19   & 7.0  & 11.59\\
                             & 10  & 4.45   & 5.8  & 23.56\\
                             & 100 & 0.497  & 0.27 & $\sim 100\%$\\
  \end{tabular}
  \caption{Number of events populating the three signal regions of the
    CMS-EXO-16-022 analysis for the different considered stop decay lengths.
    We compare the CMS and \ma~(MA5) results in the second
    and third column of the table, respectively, and evaluate the difference
    according to Eq.~\eqref{eq:5-diff} in the last column of the table.}
  \label{Analysis_compare}
\end{center}
\end{table}

In Table.~\ref{Analysis_compare}, we compare our predictions (MA5) with the
official results provided by CMS, for the four considered stop lifetimes.
The deviations are evaluated relatively to the CMS official results, according
to the measure
\be\label{eq:5-diff}
  |{\rm error}| = 
    \left|\frac{{\rm MA5}-{\rm CMS}}{{\rm CMS}}\right| \ .
\ee
We obtain a good agreement in most of the case, with the exception of the very
long stop lifetime setup ($c\tau=100$~cm) for which very important discrepencies
are found. The origins of the discrepencies are connected to the reconstruction
and selection efficiencies of Figure~\ref{fig:5-efficiencies} that have been
extracted from 8 TeV data and provided for stop decays lengths of at most
2.2~cm. More information would be necessary to allowing for better modeling
of the reconstruction properties of very long-lived stops, as we manually set
the efficiency to zero in our {\sc Delphes} configuration card. Moreover, the
position of the secondary vertex along the collision axis is used in the
CMS-EXO-16-022 analysis, so that the dependence of the efficiencies on the
longitudinal impact parameter may be important.

\section{Summary}
The \ma\ implementation of the CMS-EXO-2016-22 analysis, a search for long-lived
particles decaying into electrons and muons, has been presented. The simulation
of signal events needs to be performed using a special tune of {\sc Delphes 3}
that has been modified for handling displaced vertex information. A link to a
download of this tune is made available on the webpage\\
\hspace*{0.5cm} \url{
  https://madanalysis.irmp.ucl.ac.be/wiki/MA5LongLivedParticle}.\\
For the considered benchmark scenarios, the calculation of the signal acceptance
and efficiency is consistent with predictions given by CMS for proper decay
lengths smaller than 10 cm. However, this implementation is not valid and should
not be used to constrain models containing particles with proper decay lengths
greater than 10 cm.
This analysis is thus considered as validated and has been made
available from the \ma\ Public Analysis Database and from
{\sc InSpire}~\cite{1667603},\\
\hspace*{0.7cm}\url{http://doi.org/10.7484/INSPIREHEP.DATA.UFU4.99E3}.

\section*{Acknowledgment}
JC is in particular grateful to the CMS exotica conveners and Jamie Antonelli
for the enlightening discussions. JC also thanks Samuel Bein, Eric~Conte and
Jory Sonneveld for useful advices and good tutoring on
\textsc{ MadAnalysis5} and event simulation, as well as Dayoung Kang, Peiwen Wu
and Seungjin Yang for useful discussions.

%% file: 4-cms_sus_16_041.tex
\chapter{CMS-SUS-16-041: a CMS supersymmetry search with multileptons and jets
(35.9~fb$^{-1}$)}
\label{cms-sus-16-041}
{\it G.~Chalons, B.~Fuks, K.~Lee, J.~Park}

\begin{abstract}
We summarize the implementation within the \ma\ framework of the CMS search for
new physics through final-state signatures comprised of a least three leptons
(electrons or muons), jets and missing transverse energy. This analysis uses
$35.9$~fb$^{-1}$ of data collected in 2016 in
proton-proton collisions at a center-of-mass energy of $\sqrt{s}=13$~TeV. We
validate our implementation by comparing our results against cutflows
provided on the official CMS analysis webpage for well-defined benchmnark
scenarios.
\end{abstract}

\section{Introduction}
\begin{figure}[t]
\centering
  \includegraphics[width=0.48\linewidth]{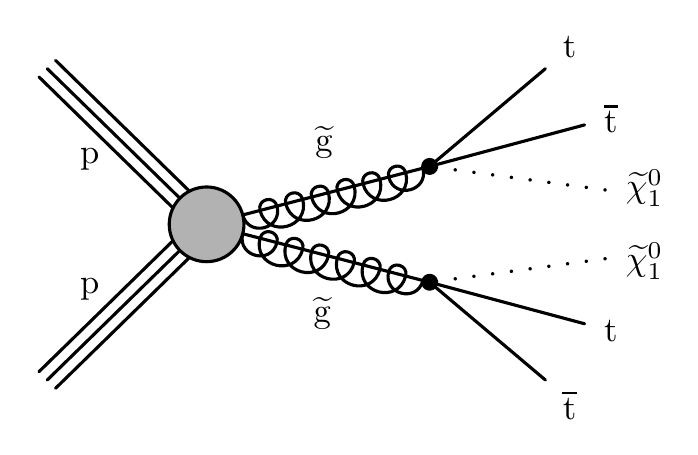}
  \includegraphics[width=0.48\linewidth]{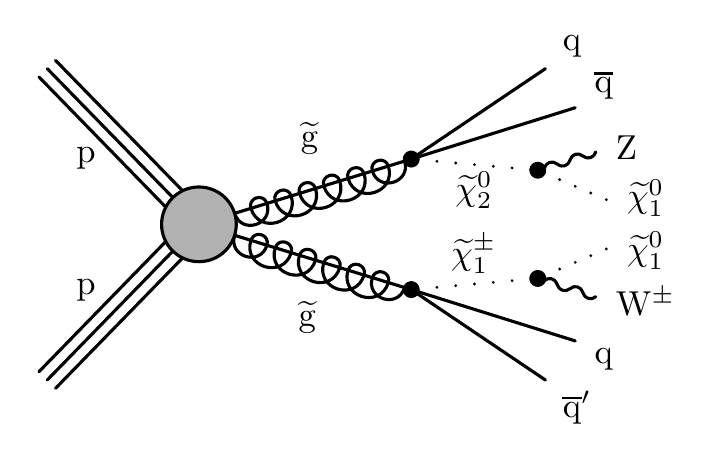}
  \caption{Representative Feynan diagrams for the two processes on which our
  reimplementation of the CMS-SUS-16-041 search has been valided. A pair of
  gluinos is produced and further decays into four top-quarks and missing energy
  (left) or into jets, missing energy and weak bosons via intermediate weak
  bosons (right).}
  \label{fig:diagram}
\end{figure}

Many models of new physics beyond the Standard Model predict processes leading
to the production of multileptonic systems. In a recent supersymmetry analysis
of 35.9~fb$^{-1}$ of proton-proton collisions at a
center-of-mass energy of $\sqrt{s}=13$ TeV~\cite{Sirunyan:2017hvp}, the CMS
collaboration has scrutinized multileptonic events in which the final state also
contains jets and some missing transverse energy. In this note, we summarize the
implementation in the \ma\ framework~\cite{Conte:2012fm,Conte:2014zja,%
Dumont:2014tja} of this search, and we describe its validation. The latter
focuses on two supersymmetric signals in
which pairs of gluinos are produced, and where each gluino decays either into a
system made of a $t \bar{t}$ pair and the lightest supersymmetric particle
(taken to be a neutralino $\tilde\chi_1^0$) that leaves the detector invisibly,
or into a pair of quarks and a heavier neutralino  $\tilde\chi_2^0$ and a
chargino $\tilde{\chi}_1^{\pm}$ that further decay
into a $Z$-boson and a $W$-boson, respectively. These two processes are
illustrated through representative Feynman diagrams in Fig.~\ref{fig:diagram}.

\section{Description of the analysis}

The analysis preselects events containing at least three leptons (electrons or
muons) and at least two jets, after having reconstructed the final-state physics
objects.

More precisely, jets are reconstructed by using the anti-$k_T$
algorithm~\cite{Cacciari:2008gp} with a radius parameter set to $R=0.4$, and
only those with a transverse momentum $p_T^j$ and pseudorapidity $\eta^j$
satisfying
\be
  p_T^j > 30~{\rm GeV} \qquad\text{and}\qquad |\eta|^j < 2.4
\ee
are retained. Jets are identified as $b$-jets by relying on the CMS cMVAv2
algorithm with its medium working point~\cite{CMS:2016kkf}, which corresponds
to a typical tagging efficiency of 70\% for a mistagging rate of charmed and
lighter jets of 10\% and 1\%, respectively. Our reimplementation of the fitted
$b$-tagging efficiency and mistagging rate provided by CMS in Table~2 of
Ref.~\cite{CMS:2016kkf} includes a global rescaling factor of 0.94 to account
for the drop in efficiency that has been observed at the time of data-taking,
in 2015-2016.

In addition, only muons and electrons with respective
pseudorapidities $\eta^e$ and $\eta^\mu$ satisfying
\be
 |\eta^e| < 2.5 \qquad\text{and}\qquad
 |\eta^\mu| < 2.5
\ee
are considered. Moreover, to discriminate leptons originating from the decays of
$W$-bosons and $Z$-bosons from those issued from hadron decays or misidentified
jets as leptons, an additional requirement on the lepton isolation is enforced
by using three different variables. The first variable is the lepton relative
isolation $I_{\rm mini}$ defined as the ratio between the amount of measured
energy in a cone of radius $\Delta R$ centered around the lepton direction and
the lepton $p_T$, with
\be
 \Delta R = \frac{\mbox{10 GeV}}{\mbox{min}(\mbox{max}(p_T(\ell),50),200)} \ .
\ee
The next two variables are computed on the basis of the lepton momentum and the
momentum of the jet that is geometrically matched to the lepton. This jet is the
jet of transverse momentum larger than 5~GeV that is the closest, in the
transverse plane, to the lepton.
The second employed variable then consists in the ratio between the lepton $p_T$
and the $p_T$ of this jet,
\begin{equation}
 p_T^{\rm ratio} = p_T(\ell) / p_T(\mbox{jet}) \ ,
\end{equation}
and the last variable is the relative lepton transverse momentum $p_T^{\rm rel}$
defined as the magnitude of the component of the lepton momentum perpendicular
to the axis of this jet. A lepton is then considered as isolated if
\begin{equation}
 I_{\rm mini}  < I_1\qquad\text{and}\qquad  \bigg[(p_T^{\rm ratio} > I_2)
  \ \ \text{or}\ \  (p_T^{\rm rel} > I_3) \bigg] \ .
\end{equation}
For muons (electrons), the selection requirements are fixed to $I_1 = 0.16$
(0.12), $I_2 = 0.69$ (0.76) and $I_3 = 6.0$~GeV (7.2~GeV) whilst loosely
isolated leptons consist of lepton candidates only fullfilling
$I_{\rm mini} < 0.4$.

The preselected events are then classified according to the value of the
hadronic transverse energy
\be
   H_T = \sum_{\rm jets} p_T\  ,
\ee
when only jets with a $p_T$ larger than 30~GeV are included in the sum.
Requirements are
finally imposed on the transverse momentum of the leading lepton $\ell_1$
and of the next-to-leading lepton $\ell_2$, depending on
the $H_T$ value.
\be
 \left\{\begin{array}{l}
    H_T < 300~{\rm GeV}: \qquad p_T(\ell_1) > 25~{\rm GeV}\ , \
      p_T(\ell_2) > x~{\rm GeV}\\
    H_T > 300~{\rm GeV}: \qquad
      p_T(\ell_1, \ell_2) > x~{\rm GeV}
 \end{array}\right. \ ,\ee
where $x=10$~GeV and 15~GeV for muons and electrons respectively. In addition,
the third lepton transverse momentum is required to satisfy
\be
  p_T(\ell_3) > 10~{\rm GeV} \ .
\ee
Moreover, the invariant mass of any pair of opposite-charge same-flavor leptons
is required to be larger than 12~GeV,
\be
  m_{\ell\ell} > 12~{\rm GeV}.
\ee
The baseline selection finally requires an amount of missing energy
\be
  E_T^{\rm miss} > 50~{\rm GeV} \quad\text{or}\quad 70~{\rm GeV} \ ,
\ee
the second requirements being only relevant for regions exhibiting a number of
$b$-jets of at most one and an $H_T$ value smaller than 400~GeV.

\begin{figure}[t]
\centering
 \includegraphics[width=\textwidth]{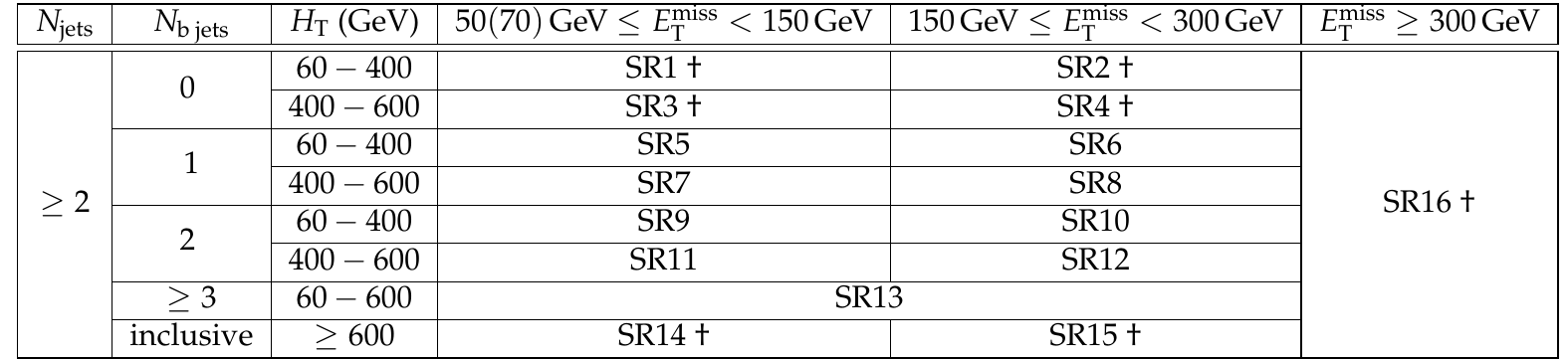}
 \caption{\label{SRs} Definition of the signal regions. The dagger indicates the
   signal regions that are further subdivided according to the value of the
   transverse mass of the system made of the missing transverse momentum and the
   lepton no connected to the $Z$-boson.}
 \includegraphics[width=\textwidth]{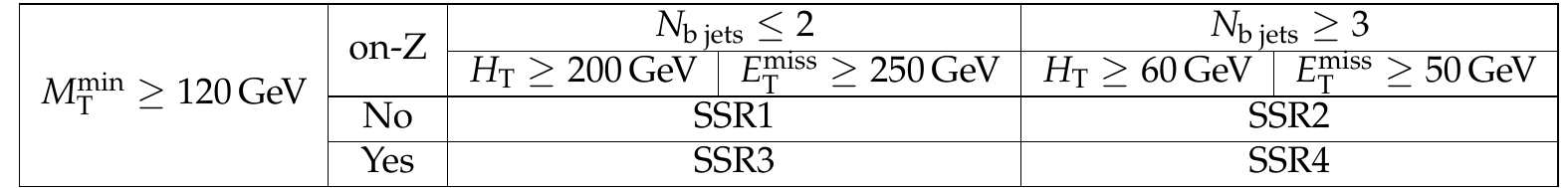}
  \caption{\label{SSRs} Definition of the aggregated signal regions.}
\end{figure}

The events are then classified into varied signal regions according to the
number of identified $b$-jets, the amount of missing transverse momentum and
the actual $H_T$ value, as summarized in Fig.~\ref{SRs} (usual signal regions)
and Fig.~\ref{SSRs} (super, or aggregated, signal regions). In addition, each
region is further divided into two regions, depending whether an opposite-sign
same-flavor lepton pair has an invariant-mass compatible with the $Z$-boson
mass, $|m_{\ell\ell}-M_Z| < 15$~GeV (on-Z) or not (off-Z), and some regions
include an requirement on the transverse mass of the system made of the missing
transverse momentum and the third lepton ($M_T < 120$~GeV or $M_T> 120$~GeV).

\section{Validation}
For the validation of our implementation, two cutflow tables have been provided
in Ref.~\cite{Sirunyan:2017hvp}. The first one concerns gluino pair production
with four top quarks in the final state, assuming gluino and neutralino masses
equal to 1500~GeV and 200 GeV respectively. The second cutflow also concerns
gluino
pair production, but in a configuration in which the gluinos decay into two
weak vector bosons and light jets (as well as missing energy) and where the
gluino and neutralino masses are fixed to 1200~GeV and 400~GeV, respectively.
For both
scenarios, the branching fraction of the gluino into (top or lighter) quarks are
assumed to be 2/3 for $\tilde{g} \to
\chargo q \bar{q}'$ and $\tilde{g} \to \neutt q \bar{q}$ with $m_{\chargo} =
m_{\neutt} = (\mneuto + m_{\tilde{g}})/2$.

\subsection{Event generation}
\label{sec:events041}
Our validation procedure includes the simulation of hard-scattering events for
the signal process
\be
  p\,p \rightarrow \tilde{g}\, \tilde{g}\ .
\ee
We have made use of the \textsc{MadGraph5\_ aMC@NLO} program
version~2.6.0~\cite{Alwall:2014hca} to simulate 10000 signal events at the
leading-order accuracy in QCD, relying for the hard process on the
SLHA2~\cite{Skands:2003cj,Allanach:2008qq} implementation of the MSSM in
MG5\_aMC~\cite{Duhr:2011se}. All
superpartners but the gluino and the lightest neutralinos and charginos have
been decoupled, their mass being set to $10^5$~GeV. The hard
matrix-element has been convoluted with the \texttt{NNPDF30\_lo\_as\_0130} set
of parton densities~\cite{Ball:2014uwa} accessed through the LHAPDF~6
library~\cite{Buckley:2014ana}.
In addition to the above process, we have also generated events for gluino pair
production in association with one and two extra
jets. Parton showering and hadronization have then been simulated by employing
the {\sc Pythia 8.260} package~\cite{Sjostrand:2014zea} with the {\sc CUETP8M1}
tune~\cite{Khachatryan:2015pea} and standard CMS settings for the matching
parameters and {\sc Pythia}~8 common settings. The merging of the multipartonic
matrix elements is performed through the MLM scheme~\cite{Mangano:2006rw}, by
imposing a minimum jet measure $k_T$ larger than 30~GeV and a merging scale of
42~GeV.

\begin{figure}
\centering
  \includegraphics[width=0.48\linewidth]{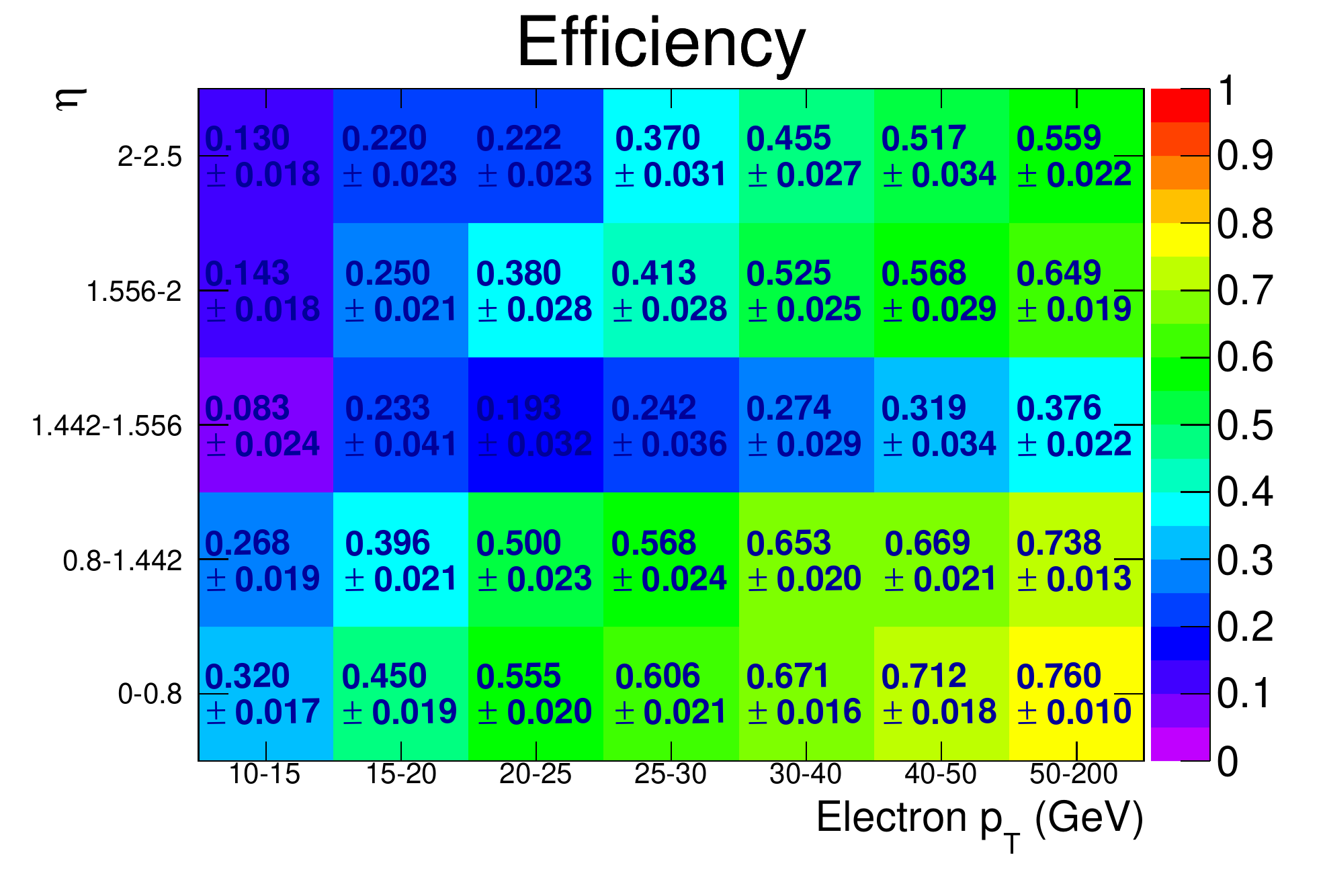}
  \includegraphics[width=0.48\linewidth]{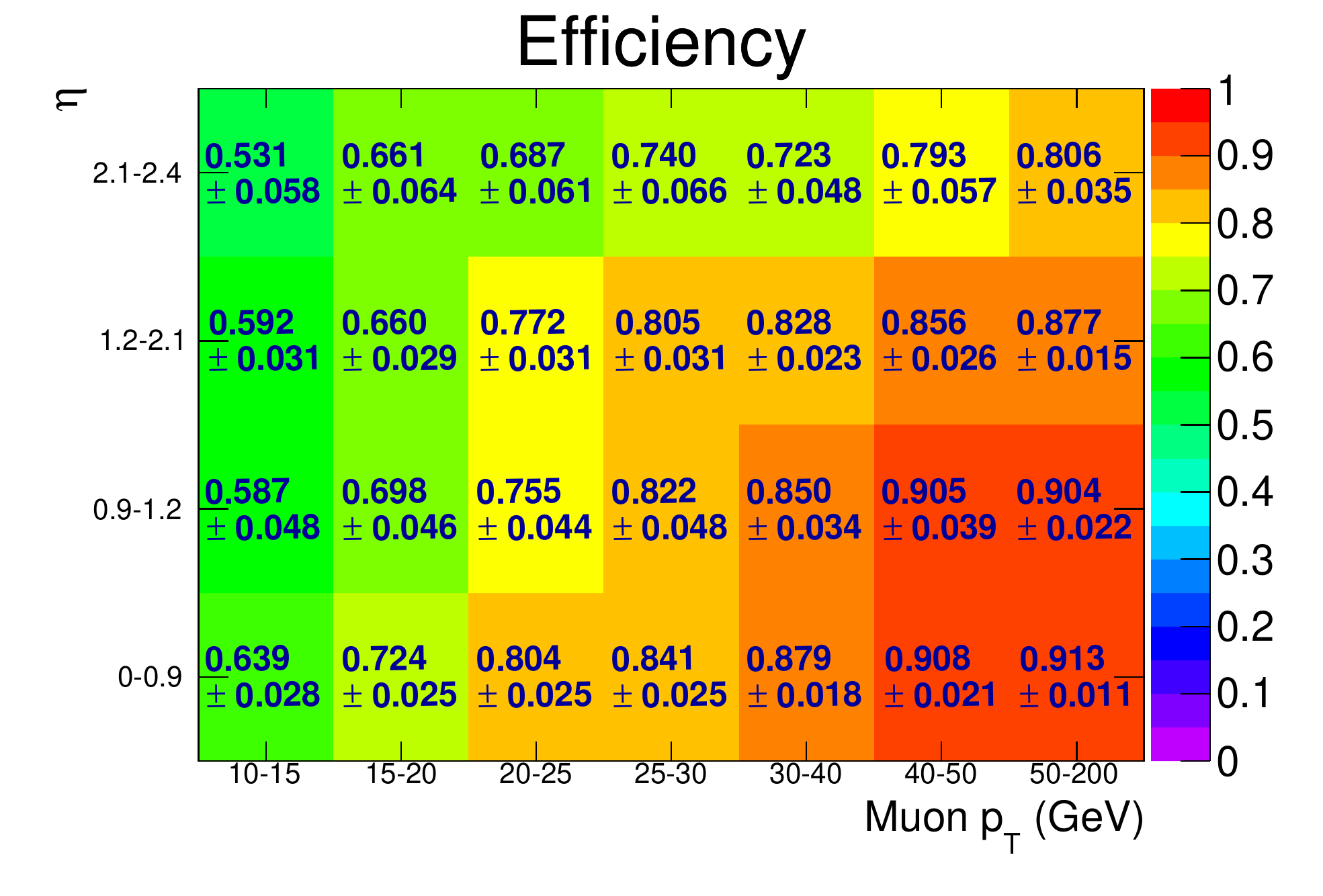}
  \caption{Representative electron (left) and muon (right) reconstruction
    efficiencies used in the CMS-SUS-16-041 analysis. The results include the
    dependence of the efficiencies on the transverse momentum $p_T$ and the
    pseudorapidity $\eta$.}
\label{fig:lepton_efficiencies}
\end{figure}

We have simulated the response of the CMS detector with the \textsc{Delphes
v3.4.1} program~\cite{deFavereau:2013fsa}, that
internally relies on {\sc FastJet}~\cite{Cacciari:2011ma} for object
reconstruction, after relaxing all isolation requirements in the {\sc Delphes}
configuration card so that isolation could be imposed at the analysis level.
We have used the $b$-tagging performances presented in Ref.~\cite{CMS:2016kkf},
although we have additionally included an overall rescaling factor of 0.94. Our
analysis uses the medium working point (see Table~2 in Ref.~\cite{CMS:2016kkf}).
We have additionally made use of the updated lepton reconstruction efficiencies
presented in Ref.~\cite{efficiency} and illustrated in
Fig.~\ref{fig:lepton_efficiencies}.

\subsection{Comparison with the official results}

\begin{table}
 \renewcommand{\arraystretch}{1.4}
  \centering
  \begin{tabular}{c|cc|cc|c}
  Selection & CMS & Efficiency (\%) & MA5 & Efficiency (\%) & Difference (\%) \\
\hline 
No selection & 509.0 & 100\% & 27345 & 100 & 0 \\ 
Trigger ($\ge$ 3 leptons) & 6.7 & 1.32 & 348 & 1.27 & -3.79 \\ 
$\ge$ 2 jets & 6.7 & 1.32 & 342 & 1.25 & -5.30 \\ 
$p_T^{\mathrm{miss}} > 50$ GeV & 6.7 & 1.32 & 337 & 1.23 & -6.82 \\ 
off-Z SR & 6.0 & 1.18 & 302 & 1.10 & -6.78 \\ 
\hline 
off-Z SR16a & 1.8 & 0.35 & 93 & 0.34 & -2.86 \\  
off-Z SR16b & 2.5 & 0.49 & 133 & 0.49 & 0 \\ 
\end{tabular}
\caption{\label{4t_cutflow} Comparison of the cutflow predicted by {\sc
  MadAnalysis 5} with the one provided by CMS for the benchmark scenario in
  which gluinos decay into top quarks and missing energy. In the last column, we
  evaluate the agreement between the results relatively to the CMS ones, as
  given in Eq.~\eqref{eq:error}.}

\vspace{0.5cm}

\begin{tabular}{c|cc|cc|c}
Selection & CMS & Efficiency (\%) & MA5 & Efficiency (\%) & Difference (\%) \\ 
\hline 
No selection & 3072.0 & 100\% & 25481 & 100 & 0 \\ 
Trigger ($\ge$ 3 leptons) & 9.6 & 0.31 & 78 & 0.31 & 0 \\ 
$\ge$ 2 jets & 9.6 & 0.31 & 78 & 0.31 & 0 \\ 
$p_T^{\mathrm{miss}} > 50$ GeV & 9.5 & 0.31 & 77 & 0.30 & -1.00 \\ 
on-Z SR & 9.1 & 0.30 & 69 & 0.27 & -3.00 \\ 
\hline 
on-Z SR15b & 1.3 & 0.04 & 15 & 0.06 & +50.00 \\  
on-Z SR16b & 5.2 & 0.17 & 34 & 0.13 & -23.53 \\ 
\end{tabular}
\caption{\label{4q_cutflow} Same as in Table~\ref{4t_cutflow} but for the
  benchmark scenario in which the gluino decays into light jets and gauge
  bosons.}
\end{table}

The provided validation material only included cutflow tables for two
well-defined benchmark scenarios, as above-mentioned. In this section, we
compare predictions obtained with {\sc MadAnalysis 5} (MA5) (and the simulation
chain introduced in Section~\ref{sec:events041}) with official CMS numbers.
Results for the gluino decays into top quarks are shown in
Table~\ref{4t_cutflow} and into lighter quarks and vector bosons in
Table~\ref{4q_cutflow}. We observe a generally good agreement, all efficiencies
being consistent with each other, except for the on-$Z$ signal regions where a
$Z$-boson is reconstructed. In this case, deviations of 30\%--50\% are obtained,
and they point either to the definition of the transverse variables used in the
analysis, or to statistics. Unfortunately, the absence of any public release of
additional pieces of information by CMS prevents us from further investigating
the issue.

\section{Conclusion}
In this chapter, we have reimplemented, in the {\sc MadAnalysis} 5 framework, a
CMS search for supersymmetry in a final state made of several leptons and jets.
The analysis focuses on a signatures constituted of a least three leptons
(electrons or muons) and uses $35.9\, \mbox{fb}^{-1}$ of data collected in
2016 at a center-of-mass energy of $\sqrt{s}=13$~TeV \cite{Sirunyan:2017hvp}.
Whilst it only contains four signal regions (off-Z SR16a, off-Z SR16b,
on-Z SR15b and on-Z SR16b) for which CMS provided cutflow tables for validating
our reimplementation~\cite{cutflows}, all the signal regions have been
implemented in our code. Whilst one of the considered benchmark scenario, in
which a gluino decays into top quarks and missing energy, provide a very good
agreement when comparing our predictions with CMS results, large discrepancies
of 30\%--50\$ have been observed for the second considered benchmark in which
the gluino decays into a gaugo boson, light jets and missing energy. The
information provided by CMS has not allowed us to further investigate the
origins of the discrepancies.

This analysis being far from being validated as a result of a lack of
information from CMS allowing to understand the source of the differences
between the CMS results and the \ma\ predictions, it has not been included in
\ma.

\section*{Acknowledgement}
The authors thank the CMS SUSY conveners for their help, and in particular
Claudio Campagnari and Lesya Shchutska who were invaluable to achieve this work.

%% file: 6-cms_sus_17_001.tex
\chapter{CMS-SUS-17-001: a CMS search for stops and dark matter with
  opposite-sign dileptons}
\label{cms-sus-17-001}
{\it S.~Bein, S.-M.~Choi, B.~Fuks, S.~Jeong, D.-W.~Kang, J.~Li, J.~Sonneveld}

\begin{abstract}
We present the \ma\ implementation and validation of the CMS-SUS-17-001
analysis, which documents a search for the production of top squarks decaying
into a dileptonic system and missing transverse energy. The results are based on
a dataset
of proton-proton collisions recorded by CMS with a center-of-mass energy of
13~TeV and an integrated luminosity of 35.9~fb$^{-1}$. The validation of our
reimplementation is based on a comparison of the expected number of signal event
counts in the signal regions with information provided by the CMS collaboration,
with signal events corresponding to a simplified scenario in which the Standard
Model is extended by a stop and a neutralino.
\end{abstract}

\section{Introduction}
In this contribution, we present the \ma~\cite{Conte:2012fm,Conte:2014zja,%
Dumont:2014tja} implementation of the CMS-SUS-17-001
search~\cite{Sirunyan:2017leh} for the superpartners
of the top quark, together with its validation. The CMS analysis targets the
production of a pair of top squarks that decay into a final-state system
comprising at least two jets with one of them being $b$-tagged, one pair of
leptons of opposite electric charge, and a significant amount
of missing transverse momentum. The main search variable consists of the $m_{T2}$ stransverse
mass~\cite{Lester:1999tx,Cheng:2008hk} that has a kinematic endpoint for the
dominant contributions to the Standard Model background.

In order to validate our reimplementation, we have reinterpreted the results of
the CMS collaboration in the context of a class of simplified models where the
Standard Model is supplemented by a top squark and a neutralino, where the neutralino is stable
and thus gives rise to missing transverse momentum. We have compared, for two benchmark
configurations, predictions obtained with our \ma\ reimplementation with the
official CMS results at different level of the selection strategy. Although the
analysis is also sensitive to generic dark matter simplified models, the
information provided by CMS has not allowed us to generate events to perform a
comparison in this case.

\section{Description of the analysis}
\begin{figure}[t]
  \centering
    \includegraphics[width=0.30\textwidth]{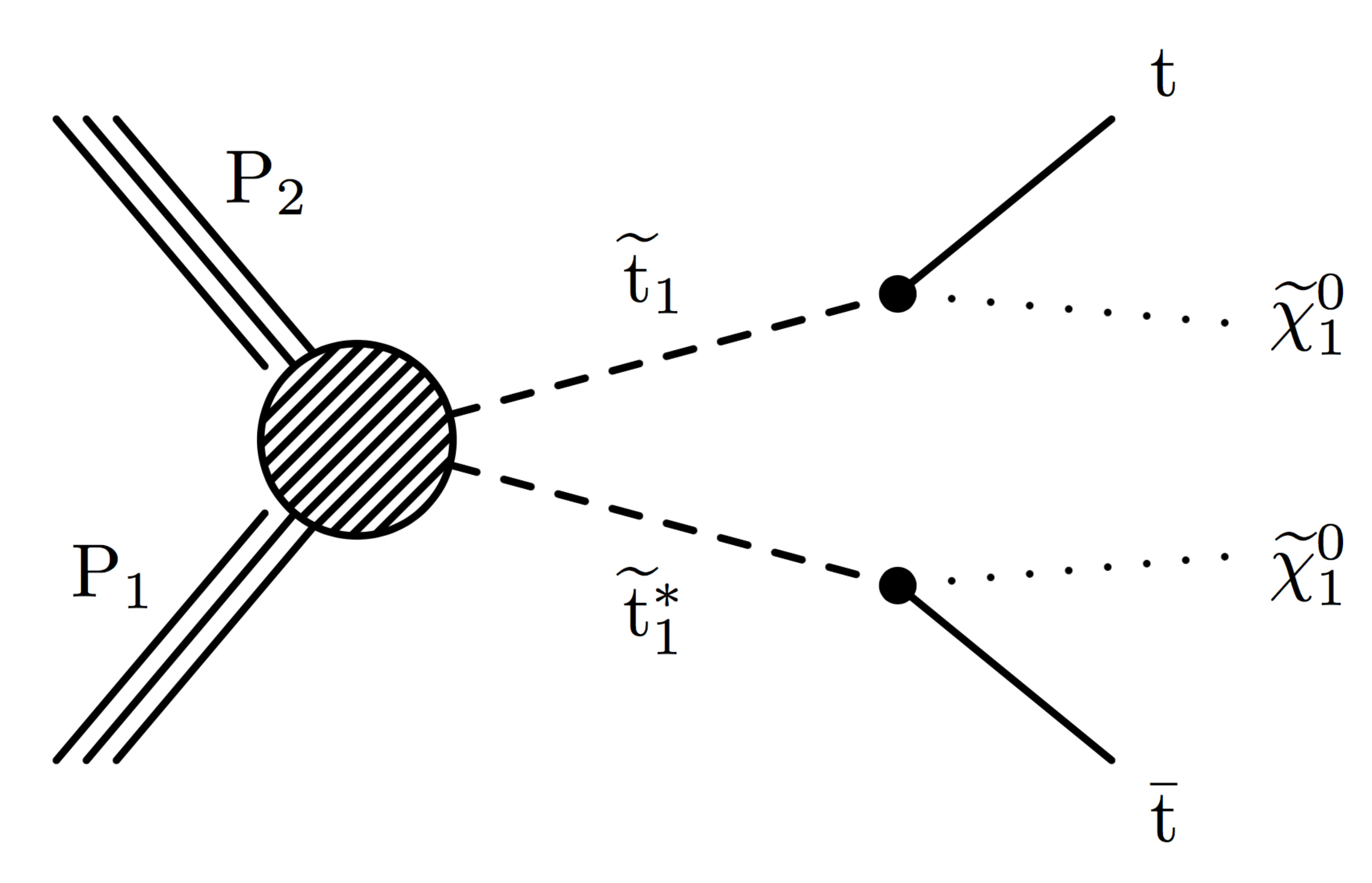}
  \caption{Representative Feynman diagram for the production of a pair of top
  squarks that each decays into a neutralino and a top quark.}
  \label{fig:6-diagrams}
\end{figure}
The CMS-SUS-17-001 analysis relies on a final-state signature made of two top
quarks and missing transverse energy $\slashed{E}_T$ as could arise from
stop-pair production and decay,
\be
  p p \to \tilde t\tilde t^* \to t \bar t + \slashed{E}_T \ ,
\ee
and illustrated in Fig.~\ref{fig:6-diagrams}
The analysis focuses on the dileptonic decay of the top-antitop system and the
preselection is implemented accordingly.

\subsection{Object definitions and preselection}
The signal region definitions rely on the presence of two lepton candidates
$\ell_1$ and $\ell_2$ whose transverse momentum $p_T$ and pseudorapidity $\eta$
satisfy
\be\bsp
  & p_T^{\ell_1} >25~{\rm GeV} \qquad\text{and}\qquad |\eta^{\ell_1}|<2.4\ ,\\
  & p_T^{\ell_2} >20~{\rm GeV} \qquad\text{and}\qquad |\eta^{\ell_2}|<2.4\ ,
\esp\ee
for the leading and next-to-leading lepton, respectively. Lepton isolation is
enforced by requiring that the sum of the transverse momentum of the particles
present in a cone of radius $R=0.3$ centered on the lepton is smaller that 0.12
times the lepton $p_T$,
\be
  \frac{1}{p_T^\ell}\sum_{i} (p_{T})_i < 0.12\ .
\label{eq:6-leptiso}\ee
Jets are recontructed by means of the anti-$k_T$
algorithm~\cite{Cacciari:2008gp} with a radius parameter set to $R=0.4$, and
their transverse momentum $p_T^j$ and pseudorapidity $\eta^j$ are required to
fulfill
\be
  p_T^j>30~{\rm GeV} \qquad\text{and}\qquad |\eta|<2.45 \ .
\ee
Moreover, any jet found within a cone of radius $R=0.4$ centered on an isolated
lepton is removed from the jet collection. Jets are tagged as $b$-jets according
to the medium working point of the CSVv2 CMS
algorithm~\cite{Chatrchyan:2012jua}, which
corresponds to a tagging efficiency of about 55\%--65\% for a percent-level
mistagging rate. The missing transverse momentum
${\bf E}_T^{\rm miss}$ is defined as the negative of the vector sum of the
transverse momenta of all reconstructed objects, and the missing transverse energy is then
defined by its norm,
\be
  E_T^{\rm miss}=|{\bf E}_T^{\rm miss}|\ .
\ee

Event preselection starts by requiring an opposite-charge
pair of leptons (electrons or muons) with a dilepton invariant mass
$m_{\ell\ell}$ satisfying
\be
  m_{\ell\ell} > 20~{\rm GeV}.
\ee
Moreover, events featuring a third loosely isolated lepton with a transverse
momentum larger than 15~GeV are vetoed. Loose lepton isolation is defined as in
Eq.~\eqref{eq:6-leptiso}, but with a different threshold,
\be
  \frac{1}{p_T^\ell}\sum_{i} (p_{T})_i < 0.40\ .
\ee
In order to suppress the Drell-Yan background, the dilepton system cannot be
compatible with a $Z$-boson and its invariant mass has to satisfy
\be
  \big| m_{\ell\ell}-m_Z\big| > 15~{\rm GeV},
\ee
when the lepton flavors are identical. To further suppress boson production
backgrounds, the analysis requires at least two jets, with at least
one of them being $b$-tagged,
\be
  N_j \geq 2 \qquad\text{and}\qquad N_b\geq 1 \ ,
\ee
where $N_j$ and $N_b$ respectively indicate the number of jets and $b$-tagged
jets. Finally, the missing transverse momentum is imposed to fulfill
\be
  E_T^{\rm miss} > 80~{\rm GeV}
  \qquad\text{and}\qquad
  S \equiv \frac{E_T^{\rm miss}}{\sqrt{H_T}} > 5~{\rm GeV}^{1/2}\ ,
\ee
the hadronic activity $H_T$ being defined as the scalar sum of the transverse
momentum of all reconstructed jets. The missing momentum is finally enforced to
be well separated in azimuth from the two leading jets $j_1$ and $j_2$,
\be
  c_1\equiv \cos\Delta\phi\big({\bf E}_T^{\rm miss}, j_1\big) < 0.80
  \qquad\text{and}\qquad
  c_2\equiv \cos\Delta\phi\big({\bf E}_T^{\rm miss}, j_2\big) < 0.96 \ .
\ee

\subsection{Event Selection}
Our implementation includes all three aggregated signal regions defined in the
CMS-SUS-17-001 analysis. Each signal region is defined by a different selection
on the amount of missing transverse momentum $E_T^{\rm miss}$ and the value of
the
stransverse mass $m_{T2}(\ell_1\ell_2)$ evaluated by considering the visible
branches of the event to be the two leptons,
\be
  m_{T2}(\ell_1\ell_2) =
  \min_{{\bf E}_{T1}^{\rm miss}+{\bf E}_{T2}^{\rm miss}={\bf E}_T^{\rm miss}}
  \Big[ \max\big[ m_T({\bf p}_T^{\ell_1}, {\bf E}_{T1}^{\rm miss}),
                  m_T({\bf p}_T^{\ell_2}, {\bf E}_{T2}^{\rm miss})\big]\Big] \ .
\ee
Here, the minimization is made by considering all possible splittings of the
missing momentum along the two decay chains. The three signal regions are then
defined as
\be\bsp
  {\rm SR~A0} & \qquad E_T^{\rm miss} > 200~{\rm GeV} \ , \qquad
   100~{\rm GeV} < m_{T2}(\ell_1\ell_2) < 140~{\rm GeV} \ , \\
  {\rm SR~A1} & \qquad E_T^{\rm miss} > 200~{\rm GeV} \ , \qquad
   140~{\rm GeV} < m_{T2}(\ell_1\ell_2) < 240~{\rm GeV} \ , \\
  {\rm SR~A2} & \qquad E_T^{\rm miss} >  80~{\rm GeV} \ , \qquad\quad
   m_{T2}(\ell_1\ell_2) \geq  240~{\rm GeV} \ .
\esp\ee

\section{Validation}
\subsection{Event generation}

For our validation, we adopt two simplified model benchmarks inspired by the
MSSM in which the Standard Model is extended by a stop and a neutralino, all
other new physics states being decoupled. The two
points respectively feature stop and neutralino masses of $(m_{\tilde{t}},
m_{\tilde{\chi}^0_1}) =  (750, 1)$~GeV and $(600, 300)$~GeV. The top squark is
imposed to decay into a top and a neutralino with a branching ratio of 100\%.

Events have been generated with \textsc{MadGraph5\_ aMC@NLO}~%
\cite{Alwall:2014hca} and {\sc Pythia 8}~\cite{Sjostrand:2014zea}. Samples
featuring different final-state jet multiplicities have been merged through the
MLM scheme~\cite{Mangano:2006rw,Alwall:2008qv}, the \textsc{Pythia8}
\texttt{qcut} parameter ({\it i.e.} the merging scale) being set to 187.5~GeV
and the corresponding \textsc{MadGraph} \texttt{xqcut} parameter being set to
125~GeV.
The simulation of the CMS detector is then achieved with the \textsc{Delphes}~3
program~\cite{deFavereau:2013fsa}, that relies on
{\sc FastJet}~\cite{Cacciari:2011ma} for object
reconstruction, which we configure to include a $b$-tagging efficiency of 60\%
for a $p_T$-dependent mistagging rate equal to $0.1 + 0.000038*p_T$. Our samples
have been normalized to the NLO+NLL cross sections taken from
Ref.~\cite{Borschensky:2014cia}, that respectively read 0.171~pb and 0.043~pb for
600 GeV and 750 GeV squarks.

\subsection{Comparison with the official results}

\begin{table}
  \centering
  \renewcommand{\arraystretch}{1.4}
  \setlength\tabcolsep{17pt}
  \begin{tabular}{l | c c | c c }
    \multirow{2}{*}{Cut} &
      \multicolumn{2}{c|}{$(m_{\tilde t}, m_{\tilde \chi}) = (750,1)$~GeV} &
      \multicolumn{2}{c}{$(m_{\tilde t}, m_{\tilde \chi}) =  (600,300)$~GeV}\\
    & CMS & MA5 & CMS & MA5 \\
    \hline \hline
    n(OS $\mu$ or $e$)$=2$ &  - &  -&  - & - \\
    $m_{\ell\ell}>20$ GeV & 0.99 & 0.99 & 0.99 & 0.97\\
    $|m_{Z}-m_{\ell\ell}|>15$ GeV & 0.95 & 0.94) & 0.89 &0.89\\
    $N_j\geq2$ & 0.87& 0.93) & 0.85 &0.89\\
    $N_b\geq1$ & 0.73 & 0.84) & 0.83 &0.83\\
    $E_T^{\text{miss}}>80$ GeV & 0.94 & 0.95 & 0.89 & 0.88\\
    $S>5$ $\text{GeV}^{1/2}$ & 0.98 & 0.92 & 0.96 & 0.91 \\
    $c_1 <0.80$ & 0.9 &0.97 & 0.92 &0.97\\
    $c_2 <0.96$ & 1.0 &0.96 & 1.0 &0.94\\
    $M_{T2}(\ell_1\ell_2)>140$ GeV & 0.49 & 0.42 & 0.17 & 0.16\\
   \hline
     All cuts & 0.24 & 0.25 & 0.083 & 0.075
  \end{tabular}
  \caption{Comparison of the signal acceptance times efficiencies predictions
    made by \ma\ with the CMS official numbers for two benchmark scenarios and
    on a cut-by-cut basis.}
  \label{tab:6-baseline}
\end{table}

\begin{figure}
  \centering
    \includegraphics[width=0.45\textwidth]{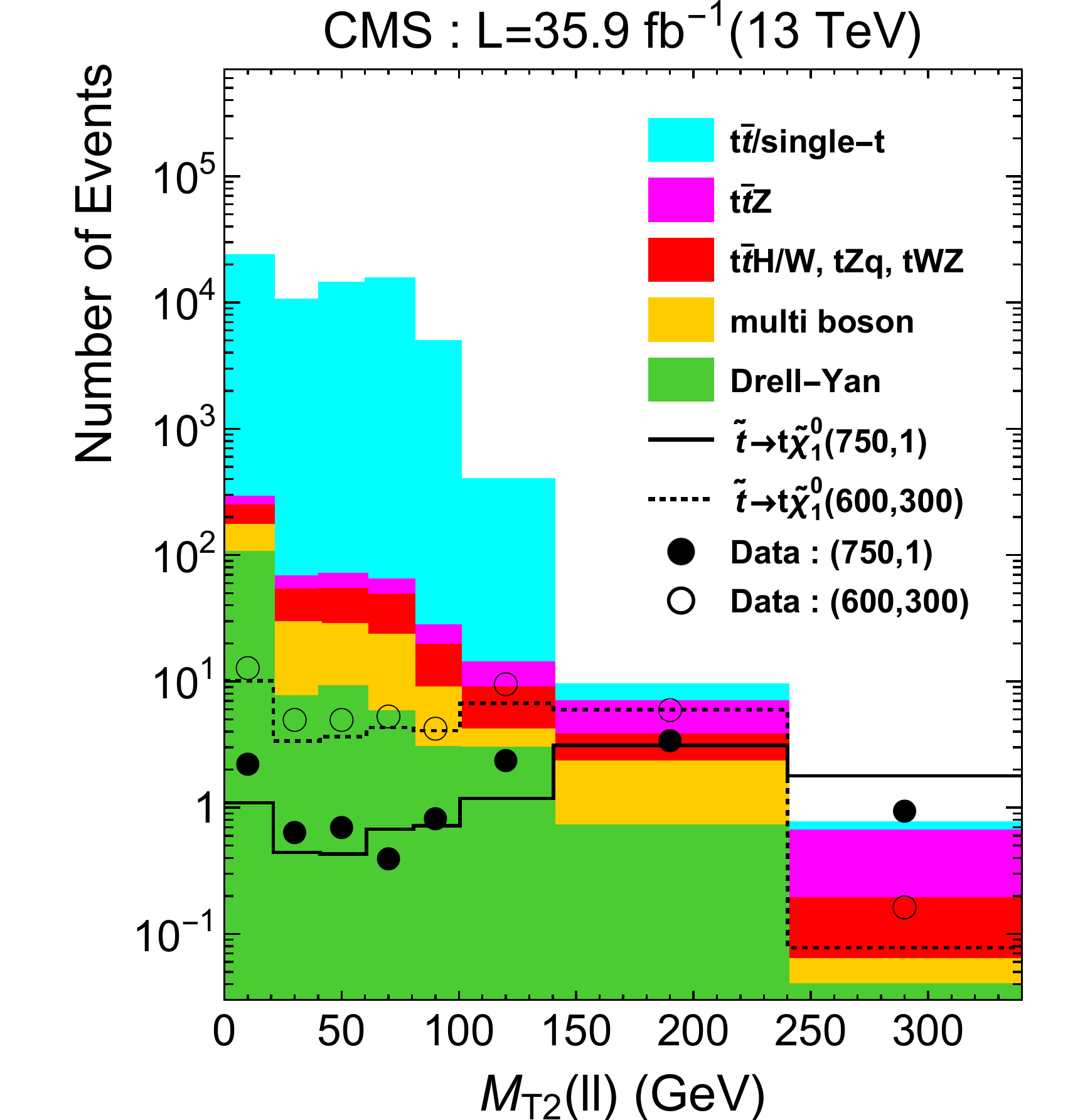}
  \caption{Comparison of the $m_{T2}$ distributions predicted by \ma\ (circles)
   with the official results provided by CMS (lines). Results for the different
   background contributions are also included (as provided by the CMS
   collaboraiton). The distributions are given after the baseline selection.}
  \label{fig:6-mt2}
\end{figure}

The cutflow for the analysis baseline selection is given in
Table~\ref{tab:6-baseline} for two supersymmetric model points of the simplified model
above-described. In this case, the $m_{T2}$ variable is required to satisfy
\be
  m_{T2}(\ell_1\ell_2) \geq  140~{\rm GeV} \ .
\ee
A comparison of the $m_{T2}(\ell_1\ell_2)$ distribution for two considered
supersymmetric scenarios is given in Fig.~\ref{fig:6-mt2}, where the lines refer
to the official CMS results and the circulars markers to the \ma\ predictions.

The final event counts of both the CMS and \ma\ results appear to agree within
20\%, similarly to the $m_{T2}$ spectra that are crucial for the final signal
region selections.

\section{Summary}
The \ma\ reimplementation of the CMS search for new physics in events with two
opposite-charge same-flavor leptons, at least one heavy-flavor tagged jet, and
large missing transverse momentum, has been presented. All baseline event
selection requirements have been incorporated, and simulated signal events for a
set of benchmark mass points have been used to validate the analysis
implementation. For the simulation, signal events corresponding to two different
choices of top squark and neutralino masses were produced in the context of the
so-called `T2tt' supersymmetric simplified model (where the Standard Model is
solely supplemented by a top squark and a neutralino) using a set of simulation
parameters synchronized with the production recipe made available by CMS.
A comparison has been made between the efficiencies of
various event selection cuts reported by CMS and the corresponding efficiencies
obtained in the \ma\ implementation, using the signal events of the
supersymmetric benchmark model points. All individual cut efficiencies, as well
as the signal efficiency after all event selection, agree within a deviation
of about 20\%. The implementation is considered to be validated, although
recasters may benefit by performing additional studies to validate the selection
efficiency in each signal region. CMS also provides correlation matrices for the
estimated background counts in the three aggregate search regions, and it may be
desirable to incorporate this information into the \ma\ implementation.
The reimplemented analysis code is available from \ma\
version 1.6 onwards, its Public Analysis Database and from {\sc InSpire}~\cite{%
1667773},\\
\hspace*{0.7cm}\url{http://doi.org/10.7484/INSPIREHEP.DATA.MMM1.876Z}.

%% file: ma5-korea-report.bbl
\providecommand{\href}[2]{#2}\begingroup\raggedright\begin{thebibliography}{10}

\bibitem{Alwall:2014hca}
J.~Alwall, R.~Frederix, S.~Frixione, V.~Hirschi, F.~Maltoni, O.~Mattelaer,
  H.~S. Shao, T.~Stelzer, P.~Torrielli, and M.~Zaro, {\it {The automated
  computation of tree-level and next-to-leading order differential cross
  sections, and their matching to parton shower simulations}},  {\em JHEP} {\bf
  07} (2014) 079, [\href{http://arxiv.org/abs/1405.0301}{{\tt
  arXiv:1405.0301}}].

\bibitem{deFavereau:2013fsa}
{\bf DELPHES 3} Collaboration, J.~de~Favereau, C.~Delaere, P.~Demin,
  A.~Giammanco, V.~Lemaître, A.~Mertens, and M.~Selvaggi, {\it {DELPHES 3, A
  modular framework for fast simulation of a generic collider experiment}},
  {\em JHEP} {\bf 02} (2014) 057, [\href{http://arxiv.org/abs/1307.6346}{{\tt
  arXiv:1307.6346}}].

\bibitem{Conte:2012fm}
E.~Conte, B.~Fuks, and G.~Serret, {\it {MadAnalysis 5, A User-Friendly
  Framework for Collider Phenomenology}},  {\em Comput. Phys. Commun.} {\bf
  184} (2013) 222--256, [\href{http://arxiv.org/abs/1206.1599}{{\tt
  arXiv:1206.1599}}].

\bibitem{Conte:2014zja}
E.~Conte, B.~Dumont, B.~Fuks, and C.~Wymant, {\it {Designing and recasting LHC
  analyses with MadAnalysis 5}},  {\em Eur. Phys. J.} {\bf C74} (2014), no.~10
  3103, [\href{http://arxiv.org/abs/1405.3982}{{\tt arXiv:1405.3982}}].

\bibitem{Dumont:2014tja}
B.~Dumont, B.~Fuks, S.~Kraml, S.~Bein, G.~Chalons, E.~Conte, S.~Kulkarni,
  D.~Sengupta, and C.~Wymant, {\it {Toward a public analysis database for LHC
  new physics searches using MADANALYSIS 5}},  {\em Eur. Phys. J.} {\bf C75}
  (2015), no.~2 56, [\href{http://arxiv.org/abs/1407.3278}{{\tt
  arXiv:1407.3278}}].

\bibitem{Aaboud:2017yqz}
{\bf ATLAS} Collaboration, M.~Aaboud et~al., {\it {Search for Dark Matter
  Produced in Association with a Higgs Boson Decaying to $b\bar b$ using 36
  fb$^{-1}$ of $pp$ collisions at $\sqrt s=13$ TeV with the ATLAS Detector}},
  {\em Phys. Rev. Lett.} {\bf 119} (2017), no.~18 181804,
  [\href{http://arxiv.org/abs/1707.01302}{{\tt arXiv:1707.01302}}].

\bibitem{Sirunyan:2017hnk}
{\bf CMS} Collaboration, A.~M. Sirunyan et~al., {\it {Search for associated
  production of dark matter with a Higgs boson decaying to
  $\mathrm{b}\bar{\mathrm{b}}$ or $\gamma \gamma$ at $ \sqrt{s}=13$ TeV}},
  {\em JHEP} {\bf 10} (2017) 180, [\href{http://arxiv.org/abs/1703.05236}{{\tt
  arXiv:1703.05236}}].

\bibitem{Aaboud:2017dor}
{\bf ATLAS} Collaboration, M.~Aaboud et~al., {\it {Search for dark matter at
  $\sqrt{s}=13$ TeV in final states containing an energetic photon and large
  missing transverse momentum with the ATLAS detector}},  {\em Eur. Phys. J.}
  {\bf C77} (2017), no.~6 393, [\href{http://arxiv.org/abs/1704.03848}{{\tt
  arXiv:1704.03848}}].

\bibitem{Aaboud:2017phn}
{\bf ATLAS} Collaboration, M.~Aaboud et~al., {\it {Search for dark matter and
  other new phenomena in events with an energetic jet and large missing
  transverse momentum using the ATLAS detector}},  {\em JHEP} {\bf 01} (2018)
  126, [\href{http://arxiv.org/abs/1711.03301}{{\tt arXiv:1711.03301}}].

\bibitem{ATLAS:2016tsc}
{\bf ATLAS} Collaboration, {\it {Search for Dark Matter production associated
  with bottom quarks with 13.3 fb$^{-1}$ of pp collisions at $\sqrt s = 13$ TeV
  with the ATLAS detector at the LHC}},  ATLAS-CONF-2016-086.

\bibitem{CMS:2016isf}
{\bf CMS} Collaboration, {\it {Search for displaced leptons in the e-mu
  channel}},  CMS-PAS-EXO-16-022.

\bibitem{Sirunyan:2017hvp}
{\bf CMS} Collaboration, A.~M. Sirunyan et~al., {\it {Search for supersymmetry
  in events with at least three electrons or muons, jets, and missing
  transverse momentum in proton-proton collisions at $\sqrt{s} = $ 13 TeV}},
  \href{http://arxiv.org/abs/1710.09154}{{\tt arXiv:1710.09154}}.

\bibitem{Sirunyan:2017leh}
{\bf CMS} Collaboration, A.~M. Sirunyan et~al., {\it {Search for top squarks
  and dark matter particles in opposite-charge dilepton final states at
  $\sqrt{s}=$ 13 TeV}},  {\em Phys. Rev.} {\bf D97} (2018), no.~3 032009,
  [\href{http://arxiv.org/abs/1711.00752}{{\tt arXiv:1711.00752}}].

\bibitem{Abercrombie:2015wmb}
D.~Abercrombie et~al., {\it {Dark Matter Benchmark Models for Early LHC Run-2
  Searches: Report of the ATLAS/CMS Dark Matter Forum}},
  \href{http://arxiv.org/abs/1507.00966}{{\tt arXiv:1507.00966}}.

\bibitem{Cacciari:2008gp}
M.~Cacciari, G.~P. Salam, and G.~Soyez, {\it {The Anti-k(t) jet clustering
  algorithm}},  {\em JHEP} {\bf 04} (2008) 063,
  [\href{http://arxiv.org/abs/0802.1189}{{\tt arXiv:0802.1189}}].

\bibitem{Degrande:2011ua}
C.~Degrande, C.~Duhr, B.~Fuks, D.~Grellscheid, O.~Mattelaer, and T.~Reiter,
  {\it {UFO - The Universal FeynRules Output}},  {\em Comput. Phys. Commun.}
  {\bf 183} (2012) 1201--1214, [\href{http://arxiv.org/abs/1108.2040}{{\tt
  arXiv:1108.2040}}].

\bibitem{Ball:2014uwa}
{\bf NNPDF} Collaboration, R.~D. Ball et~al., {\it {Parton distributions for
  the LHC Run II}},  {\em JHEP} {\bf 04} (2015) 040,
  [\href{http://arxiv.org/abs/1410.8849}{{\tt arXiv:1410.8849}}].

\bibitem{Sjostrand:2006za}
T.~Sjostrand, S.~Mrenna, and P.~Z. Skands, {\it {PYTHIA 6.4 Physics and
  Manual}},  {\em JHEP} {\bf 05} (2006) 026,
  [\href{http://arxiv.org/abs/hep-ph/0603175}{{\tt hep-ph/0603175}}].

\bibitem{Cacciari:2011ma}
M.~Cacciari, G.~P. Salam, and G.~Soyez, {\it {FastJet User Manual}},  {\em Eur.
  Phys. J.} {\bf C72} (2012) 1896, [\href{http://arxiv.org/abs/1111.6097}{{\tt
  arXiv:1111.6097}}].

\bibitem{Banerjee:2017wxi}
S.~Banerjee, D.~Barducci, G.~Bélanger, B.~Fuks, A.~Goudelis, and B.~Zaldivar,
  {\it {Cornering pseudoscalar-mediated dark matter with the LHC and
  cosmology}},  {\em JHEP} {\bf 07} (2017) 080,
  [\href{http://arxiv.org/abs/1705.02327}{{\tt arXiv:1705.02327}}].

\bibitem{1635567}
B.~Fuks and M.~Zumbihl, {\it {MadAnalysis 5 implementation of the multijet +
  missing energy analysis of ATLAS with 13.3 fb-1 of data
  (ATLAS-CONF-2016-086)}},  10.7484/INSPIREHEP.DATA.UUIF.89NC.

\bibitem{Berlin:2014cfa}
A.~Berlin, T.~Lin, and L.-T. Wang, {\it {Mono-Higgs Detection of Dark Matter at
  the LHC}},  {\em JHEP} {\bf 06} (2014) 078,
  [\href{http://arxiv.org/abs/1402.7074}{{\tt arXiv:1402.7074}}].

\bibitem{Aad:2016jkr}
{\bf ATLAS} Collaboration, G.~Aad et~al., {\it {Muon reconstruction performance
  of the ATLAS detector in proton–proton collision data at $\sqrt{s}$ =13
  TeV}},  {\em Eur. Phys. J.} {\bf C76} (2016), no.~5 292,
  [\href{http://arxiv.org/abs/1603.05598}{{\tt arXiv:1603.05598}}].

\bibitem{Aaboud:2016obm}
{\bf ATLAS} Collaboration, M.~Aaboud et~al., {\it {Search for dark matter in
  association with a Higgs boson decaying to $b$-quarks in $pp$ collisions at
  $\sqrt s=13$ TeV with the ATLAS detector}},  {\em Phys. Lett.} {\bf B765}
  (2017) 11--31, [\href{http://arxiv.org/abs/1609.04572}{{\tt
  arXiv:1609.04572}}].

\bibitem{Sjostrand:2014zea}
T.~Sjöstrand, S.~Ask, J.~R. Christiansen, R.~Corke, N.~Desai, P.~Ilten,
  S.~Mrenna, S.~Prestel, C.~O. Rasmussen, and P.~Z. Skands, {\it {An
  Introduction to PYTHIA 8.2}},  {\em Comput. Phys. Commun.} {\bf 191} (2015)
  159--177, [\href{http://arxiv.org/abs/1410.3012}{{\tt arXiv:1410.3012}}].

\bibitem{1672227}
S.~Jeon, Y.~Kang, G.~Lee, and C.~Yu, {\it {The MadAnalysis5 implementation of
  the ATLAS analysis ATLAS-EXOT-2016-25: an ATLAS mono-Higgs analysis}},
  10.7484/INSPIREHEP.DATA.SSS4.298U.

\bibitem{ATL-PHYS-PUB-2016-012}
{\bf ATLAS} Collaboration, {\it {Optimisation of the ATLAS $b$-tagging
  performance for the 2016 LHC Run}},  ATL-PHYS-PUB-2016-012.

\bibitem{Borschensky:2014cia}
C.~Borschensky, M.~Krämer, A.~Kulesza, M.~Mangano, S.~Padhi, T.~Plehn, and
  X.~Portell, {\it {Squark and gluino production cross sections in pp
  collisions at $\sqrt{s}$ = 13, 14, 33 and 100 TeV}},  {\em Eur. Phys. J.}
  {\bf C74} (2014), no.~12 3174, [\href{http://arxiv.org/abs/1407.5066}{{\tt
  arXiv:1407.5066}}].

\bibitem{Mangano:2006rw}
M.~L. Mangano, M.~Moretti, F.~Piccinini, and M.~Treccani, {\it {Matching matrix
  elements and shower evolution for top-quark production in hadronic
  collisions}},  {\em JHEP} {\bf 01} (2007) 013,
  [\href{http://arxiv.org/abs/hep-ph/0611129}{{\tt hep-ph/0611129}}].

\bibitem{Alwall:2008qv}
J.~Alwall, S.~de~Visscher, and F.~Maltoni, {\it {QCD radiation in the
  production of heavy colored particles at the LHC}},  {\em JHEP} {\bf 02}
  (2009) 017, [\href{http://arxiv.org/abs/0810.5350}{{\tt arXiv:0810.5350}}].

\bibitem{ATL-PHYS-PUB-2014-021}
{\bf ATLAS} Collaboration, {\it {ATLAS Run 1 Pythia8 tunes}},
  ATL-PHYS-PUB-2014-021.

\bibitem{1672234}
D.~Sengupta, {\it {The MadAnalysis5 implementation of the ATLAS in
  monojet+missing energy}},  10.7484/INSPIREHEP.DATA.HUH5.239F.

\bibitem{Backovic:2015soa}
M.~Backović, M.~Krämer, F.~Maltoni, A.~Martini, K.~Mawatari, and M.~Pellen,
  {\it {Higher-order QCD predictions for dark matter production at the LHC in
  simplified models with s-channel mediators}},  {\em Eur. Phys. J.} {\bf C75}
  (2015), no.~10 482, [\href{http://arxiv.org/abs/1508.05327}{{\tt
  arXiv:1508.05327}}].

\bibitem{Alloul:2013bka}
A.~Alloul, N.~D. Christensen, C.~Degrande, C.~Duhr, and B.~Fuks, {\it
  {FeynRules 2.0 - A complete toolbox for tree-level phenomenology}},  {\em
  Comput. Phys. Commun.} {\bf 185} (2014) 2250--2300,
  [\href{http://arxiv.org/abs/1310.1921}{{\tt arXiv:1310.1921}}].

\bibitem{Degrande:2014vpa}
C.~Degrande, {\it {Automatic evaluation of UV and R2 terms for beyond the
  Standard Model Lagrangians: a proof-of-principle}},  {\em Comput. Phys.
  Commun.} {\bf 197} (2015) 239--262,
  [\href{http://arxiv.org/abs/1406.3030}{{\tt arXiv:1406.3030}}].

\bibitem{1642639}
S.~Baek and T.~H. Jung, {\it {Madanalysis5 implementation of the ATLAS search
  for Dark matter in final states containing an energetic photon and large
  missing transverse momentum documented in arXiv:1704.03848}},
  10.7484/INSPIREHEP.DATA.88NC.0FER.1.

\bibitem{Baum:2017gbj}
S.~Baum, K.~Freese, N.~R. Shah, and B.~Shakya, {\it {NMSSM Higgs boson search
  strategies at the LHC and the mono-Higgs signature in particular}},  {\em
  Phys. Rev.} {\bf D95} (2017), no.~11 115036,
  [\href{http://arxiv.org/abs/1703.07800}{{\tt arXiv:1703.07800}}].

\bibitem{Khachatryan:2015iwa}
{\bf CMS} Collaboration, V.~Khachatryan et~al., {\it {Performance of Photon
  Reconstruction and Identification with the CMS Detector in Proton-Proton
  Collisions at sqrt(s) = 8 TeV}},  {\em JINST} {\bf 10} (2015), no.~08 P08010,
  [\href{http://arxiv.org/abs/1502.02702}{{\tt arXiv:1502.02702}}].

\bibitem{CMS:2016kkf}
{\bf CMS} Collaboration, {\it {Identification of b quark jets at the CMS
  Experiment in the LHC Run 2}},  CMS-PAS-BTV-15-001.

\bibitem{1642631}
S.~Ahn, J.~Park, and W.~Zhang, {\it {Madanalysis5 implementation of the CMS
  search for Dark matter with large missing transverse momentum and a Higgs
  boson decaying to a pair of photons documented in arXiv: 1703.05236}},
  10.7484/INSPIREHEP.DATA.JT56.DDC3.1.

\bibitem{Khachatryan:2014mea}
{\bf CMS} Collaboration, V.~Khachatryan et~al., {\it {Search for Displaced
  Supersymmetry in events with an electron and a muon with large impact
  parameters}},  {\em Phys. Rev. Lett.} {\bf 114} (2015), no.~6 061801,
  [\href{http://arxiv.org/abs/1409.4789}{{\tt arXiv:1409.4789}}].

\bibitem{Allanach:2002nj}
B.~C. Allanach et~al., {\it {The Snowmass points and slopes: Benchmarks for
  SUSY searches}},  {\em Eur. Phys. J.} {\bf C25} (2002) 113--123,
  [\href{http://arxiv.org/abs/hep-ph/0202233}{{\tt hep-ph/0202233}}].

\bibitem{Skands:2003cj}
P.~Z. Skands et~al., {\it {SUSY Les Houches accord: Interfacing SUSY spectrum
  calculators, decay packages, and event generators}},  {\em JHEP} {\bf 07}
  (2004) 036, [\href{http://arxiv.org/abs/hep-ph/0311123}{{\tt
  hep-ph/0311123}}].

\bibitem{1667603}
J.~Chang, {\it {Madanalysis5 implementation of CMS-EXO-16-022}},
  10.7484/INSPIREHEP.DATA.UFU4.99E3.

\bibitem{Allanach:2008qq}
B.~C. Allanach et~al., {\it {SUSY Les Houches Accord 2}},  {\em Comput. Phys.
  Commun.} {\bf 180} (2009) 8--25, [\href{http://arxiv.org/abs/0801.0045}{{\tt
  arXiv:0801.0045}}].

\bibitem{Duhr:2011se}
C.~Duhr and B.~Fuks, {\it {A superspace module for the FeynRules package}},
  {\em Comput. Phys. Commun.} {\bf 182} (2011) 2404--2426,
  [\href{http://arxiv.org/abs/1102.4191}{{\tt arXiv:1102.4191}}].

\bibitem{Buckley:2014ana}
A.~Buckley, J.~Ferrando, S.~Lloyd, K.~Nordström, B.~Page, M.~Rüfenacht,
  M.~Schönherr, and G.~Watt, {\it {LHAPDF6: parton density access in the LHC
  precision era}},  {\em Eur. Phys. J.} {\bf C75} (2015) 132,
  [\href{http://arxiv.org/abs/1412.7420}{{\tt arXiv:1412.7420}}].

\bibitem{Khachatryan:2015pea}
{\bf CMS} Collaboration, V.~Khachatryan et~al., {\it {Event generator tunes
  obtained from underlying event and multiparton scattering measurements}},
  {\em Eur. Phys. J.} {\bf C76} (2016), no.~3 155,
  [\href{http://arxiv.org/abs/1512.00815}{{\tt arXiv:1512.00815}}].

\bibitem{efficiency}
 {\em
  https://twiki.cern.ch/twiki/bin/view/CMSPublic/SUSMoriond2017ObjectsEfficiency}.

\bibitem{cutflows}
 {\em
  http://cms-results.web.cern.ch/cms-results/public-results/publications/SUS-16-041}.

\bibitem{Lester:1999tx}
C.~G. Lester and D.~J. Summers, {\it {Measuring masses of semiinvisibly
  decaying particles pair produced at hadron colliders}},  {\em Phys. Lett.}
  {\bf B463} (1999) 99--103, [\href{http://arxiv.org/abs/hep-ph/9906349}{{\tt
  hep-ph/9906349}}].

\bibitem{Cheng:2008hk}
H.-C. Cheng and Z.~Han, {\it {Minimal Kinematic Constraints and m(T2)}},  {\em
  JHEP} {\bf 12} (2008) 063, [\href{http://arxiv.org/abs/0810.5178}{{\tt
  arXiv:0810.5178}}].

\bibitem{Chatrchyan:2012jua}
{\bf CMS} Collaboration, S.~Chatrchyan et~al., {\it {Identification of b-quark
  jets with the CMS experiment}},  {\em JINST} {\bf 8} (2013) P04013,
  [\href{http://arxiv.org/abs/1211.4462}{{\tt arXiv:1211.4462}}].

\bibitem{1667773}
S.~Bein, S.-M. Choi, B.~Fuks, S.~Jeong, D.~W. Kang, J.~Li, and J.~Sonneveld,
  {\it {Madanalysis5 implementation of CMS-SUS-17-001}},
  10.7484/INSPIREHEP.DATA.MMM1.876Z.

\end{thebibliography}\endgroup
